\documentclass[aps,prb,twocolumn]{revtex4}

\usepackage{amssymb}
\usepackage{graphicx}
\usepackage{epsfig}
\usepackage{bm}
\usepackage{amsmath}
\usepackage{amsfonts}

\begin{document}

\title{$^{7}$Li NMR Study of Heavy Fermion
LiV$_{2}$O$_{4}$ Containing Magnetic Defects}
\date{\today}
\author{X. Zong}
\author{S. Das}
\author{F. Borsa}
\altaffiliation[Present address: ]{Dipartimento di Fisica ``A. Volta,''
ed Unit\`a CNISM, Universit\`a di Pavia, I-27100, Pavia,
Italy} 
\author{M. D. Vannette}
\author{R. Prozorov}
\author{J. Schmalian}
\author{D. C. Johnston}
\affiliation{Ames Laboratory and Department of Physics and Astronomy,
Iowa State University, Ames, Iowa 50011}

\begin{abstract} 
We present a systematic study of the variations of the
$^{7}$Li NMR properties versus magnetic defect concentration
$n_{\mathrm{defect}}$ within the spinel structure of polycrystalline
powder samples ($n_{\mathrm{defect}}=0.21, 0.49$, and 0.83~mol\%) and 
a collection of small single crystals ($n_{\mathrm{defect}}=0.38$~mol\%)
of LiV$_{2}$O$_{4}$ in the temperature range from 0.5 to 4.2~K\@.  We
also report static magnetization measurements and ac magnetic
susceptibility measurements at 14~MHz on the samples at low
temperatures.  Both the $^7$Li NMR spectrum and nuclear spin-lattice
relaxation rate are inhomogeneous in the presence of the magnetic
defects.  The $^7$Li NMR data for the \emph{powders} are well explained by
assuming that (i) there is a random distribution of magnetic point
defects, (ii) the same heavy Fermi liquid is present in the samples
containing the magnetic defects as in magnetically pure LiV$_{2}$O$_{4}$,
and (iii) the influences of the magnetic defects and of the Fermi liquid
on the magnetization and NMR properties are separable.  In the
\emph{single crystals}, somewhat different behaviors are observed,
possibly due to a modification of the heavy Fermi liquid, to a lack of
separability of the relaxation effects due to the Fermi liquid and
the magnetic defects, to non-Fermi liquid behavior of the conduction
electrons, and/or to quantum fluctuations of finite-size magnetic defects
(magnetic droplets).  Remarkably, the magnetic defects in the powder
samples show evidence of spin freezing below $T\approx 1.0$~K, whereas in
the single crystals with similar magnetic defect concentration no spin
freezing was found down to
$T = 0.5$~K\@.  Thus different types of magnetic defects and/or
interactions between them appear to arise in the powders versus the
crystals, possibly due to the substantially different synthesis 
conditions of the powders and crystals.
\end{abstract}


\maketitle

\section{Introduction}

LiV$_2$O$_4$ is a rare $d$-electron heavy fermion system at low
temperatures $T < 10$~K\@.\cite{Kondo1997}  The low temperature linear
electronic specific heat coefficient $\gamma$ (0.42~J/mol~K$^2$) and
Pauli magnetic susceptibility $\chi_0$ ($\approx0.01$~cm$^3$/mol) are 180
and 310 times those of a free electron gas, respectively, assuming each
vanadium atom contributes 1.5 free electrons.  The Wilson ratio
$R_{\mathrm{W}}$, which is the ratio of the enhancement factors of
$\chi_0$ and $\gamma$, is equal to 1.7, typical for a heavy fermion
system.\cite{Stewart1984}  Heavy fermion behavior was further confirmed by
electrical resistivity measurements which show a $T^2$ dependence below
2~K with a large coefficient
$A=2.2$~$\mu\Omega$~cm/K$^2$.\cite{Takagi1999,Urano2000}  The
$A$ and $\gamma$ values approximately follow the Kadowaki-Woods relation,
$A/\gamma^2=1.0\times 10^{-5}~\Omega$~cm(mol K/J)$^2$, which holds for a
variety of heavy fermion systems.\cite{Kadowaki1986}  Despite continuous
theoretical work, a detailed explanation of the heavy fermion behaviors
in LiV$_2$O$_4$ remains a
challenge.\cite{Fulde2004,Arita2007,Yushankhai2007}

$^7$Li nuclear magnetic resonance (NMR) was an important local probe in
establishing the low temperature heavy fermion behavior in magnetically
pure samples of LiV$_2$O$_4$.\cite{Kondo1997,Mahajan1998}  The low
temperature
$^7$Li nuclear spin-lattice relaxation rate $1/T_1$ follows a Korringa
relation
$1/T_1\propto T$, with a coefficient
$1/T_1T=2.2$~s$^{-1}$K$^{-1}$, which is 6000 times larger than in the
non-heavy fermion isostructural superconducting\cite{Johnston1976}
compound LiTi$_2$O$_4$.\cite{Dalton1994}  The Korringa ratio $\kappa=
[4\pi k_{\mathrm{B}}\gamma_{\mathrm{n}}^2/(\hbar 
\gamma_{\mathrm{e}}^2)]K^2T_1T$, where $K$ is the Knight shift,
$\gamma_{\mathrm{n}}$ and $\gamma_{\mathrm{e}}$ are the gyromagnetic
ratios of the $^7$Li nuclear spin and the conduction electron spin,
respectively, is equal to 0.7, which is close to the value of unity
expected for a free electron gas.

Recently, we found that the low temperature $^7$Li NMR properties of
polycrystalline LiV$_2$O$_4$ are very sensitive to the presence of a small
concentration of magnetic defects ($n_{\mathrm{defect}}=0.73$~mol\%)
within the spinel structure.\cite{Johnston2005,Kaps2001} In a
sample containing a negligible concentration of magnetic
defects, the longitudinal component of the global bulk $^7$Li nuclear
magnetization $M(t)$ after time delay $t$ following a sequence of
saturation pulses showed a single exponential recovery $1-M(t)/M(\infty) =
\exp[-(t/T_1)]$, where $1/T_1$ was proportional to $T$ as noted above. 
However, in the sample with $n_{\mathrm{defect}}=0.73$~mol\%, the $M(t)$
showed a stretched exponential recovery $1-M(t)/M(\infty) =
\exp[-(t/T_1^*)^{\beta}]$, with the characteristic relaxation rate
$1/T_1^*$ showing a peak at $T\approx0.7$~K\@.  Here 
$\beta$ is the stretching exponent with, in general, $0 < \beta < 1$. 
There was also a clear difference in the $^7$Li NMR spectrum in these two
samples.  At low temperatures $T<4.2$~K, the magnetically pure sample had
a narrow spectrum with an almost temperature independent width (full
width at half maximum peak intensity FWHM $\sim 20$~kHz). In contrast, a
strong temperature dependent inhomogeneous broadening (FWHM $\sim
100$~kHz at
$T<4.2$~K) was observed in the sample with
$n_{\mathrm{defect}}=0.73$~mol\%.

In order to further clarify the nature of the magnetic defects and their
effect on the heavy fermion properties of LiV$_2$O$_4$, we report herein 
$^7$Li NMR studies on LiV$_2$O$_4$ versus magnetic defect
concentration.  Three polycrystalline samples and a collection of single
crystals are studied. The powder samples are labeled as 6b,
7a, and 6a, with $n_{\mathrm{defect}}= 0.21$~mol\%, 0.49~mol\%, and
0.83~mol\%, respectively.  The single crystal sample is labeled as sample
1 with $n_{\mathrm{defect}}=0.38$~mol\%.  We determined the magnetic
defect concentrations from static magnetization measurements in the
temperature range 1.8--5~K and applied magnetic field range 0--5.5
T\@.\cite{Kondo1999} Furthermore, to study the possible spin freezing of
the magnetic defects at low temperatures, we measured the ac magnetic
susceptibility at 14 MHz from 0.5 to 6~K of the single crystals and of the
powder sample~6a with $n_{\mathrm{defect}}=0.83$~mol\% using the
tunnel-diode resonator technique.\cite{Vannette2007}

The temperature dependences of the $^7$Li nuclear spin-lattice relaxation
rates in our polycrystalline samples are similar to that of sample 3-3-a2
($n_{\mathrm{defect}}=0.73$~mol\%) that we studied in
Ref.~\onlinecite{Johnston2005}, which showed a peak in
$1/T_{1}^{\ast}(T)$ at about 1~K\@. However, we find a qualitative
difference in the temperature dependence of $1/T_{1}^{\ast}$ in the
collection of single crystals, which instead decreases monotonically with
decreasing temperature from 4.2~K down to 0.5~K\@. We include two
important aspects into the analysis of the NMR data.  First, we consider
the effect of a distribution of local fields due to different positions
of the $^{7}$Li nuclei relative to their nearby magnetic defects. For the
polycrystalline samples, this approach is quantitatively consistent with
the inhomogeneous broadening of the spectrum and the nonexponential
relaxation behavior. In the single crystals, this purely geometric origin
for the nuclear relaxation fails to explain the observed behavior at
$T\lesssim 1.3$~K\@.  We then extend our analysis to take into account
a possible size distribution of postulated magnetic defects of
finite size (magnetic droplets).  We speculate that the differences
between the natures and interactions of the magnetic defects/droplets in
the powders versus the crystals arise from the very different synthesis
conditions and procedures of the powders and crystals.

The paper is organized as follows. Experimental details are given in Sec.\
\ref{scn:exp}. In Sec.\ \ref{scn:results}, we report the experimental
results of the magnetization, ac susceptibility, $^{7}$Li NMR spectra,
and $^{7}$Li nuclear spin-lattice relaxation rate measurements.  In
Sec.~\ref{scn:models}, we analyze the NMR results. In
Sec.~\ref{scn:conclusions}, we summarize the main conclusions of the
paper.

\section{\label{scn:exp}Experimental Details}

Polycrystalline LiV$_{2}$O$_{4}$ samples were prepared using conventional
solid state reaction at temperatures up to 700~$^\circ$C\@.  The starting
materials were V$_2$O$_3$ (99.99\%, MV Labs),  V$_2$O$_5$ (99.99\%, MV
Labs), and Li$_2$CO$_3$ (99.999\%, Alfa Aeser).  Details of the sample
synthesis procedure can be found in Ref.~\onlinecite{Kondo1999}.  The
typical size of the polycrystalline grains is in the range of 1--10
$\mu$m,\cite{Das2006} as determined from scanning electron microscope
(SEM) micrographs.  Single crystals were grown at 950--1040~$^\circ$C
using a self-flux technique.\cite{Das2007}  The flux consisted of a
mixture of Li$_3$VO$_4$ and LiV$_{2}$O$_{4}$.  The typical size of the
crystals used in the present work is 0.2~mm.  Static magnetization
measurements were performed using a Quantum Design SQUID magnetometer in
the temperature range 1.8--350~K and applied magnetic field range
0--5.5~T\@.

The ac magnetic susceptibility was measured using a highly sensitive
self-resonating $LC$ circuit where losses are compensated by a tunnel
diode that has a region of negative differential resistance in its $I$-$V$
characteristic. The resonant frequency of an empty coil
$f_{0}=1/(2\pi\sqrt{LC})$ changes when a sample is placed in the coil.
The shift of the resonant frequency, $\Delta f=f\left( T,H\right) -f_{0}$
is directly related to the dimensionless volume ac susceptibility
$\chi_{\mathrm{ac}}\left(T,H\right)$ of the sample via\cite{Vannette2007} 
\begin{equation} {\frac{\Delta
f}{{f_{0}}}}\approx-{\frac{1}{2}}{\frac{V_{\mathrm{s}}}
{{V_{\mathrm{c}}}}}4\pi\chi_{\mathrm{ac}}, 
\label{eqn:chiac}
\end{equation} 
where $V_{\mathrm{s}}$ is the sample volume and $V_{\mathrm{c}}$ is the
coil volume.  The volume magnetization is the magnetic moment per unit
volume of the sample, with Gaussian units
G~cm$^3/\mathrm{cm}^3=\mathrm{G}$. The volume susceptibility is the
volume magnetization divided by field, which is then dimensionless. The
optimized and thermally stabilized circuit resonates at 14~MHz with a
stability of $0.05\mathrm{\ Hz}$ over hours.\cite{Vannette2007} The
resonator was mounted in a $^3$He cryostat with a temperature range
0.5--150 K\@. A static external field up to 90~kOe can be applied to
study field-dependent properties.

$^{7}$Li NMR measurements were performed utilizing a phase-coherent pulse
spectrometer at applied magnetic fields $H= 1.06$, 1.68 and 3.0~T and in
the temperature range 0.5--4.2 K\@. Measurements above 1.5 K were
performed with a $^4$He bath cryostat and measurements below 1.5 K with a
Janis $^3$He cryostat. The typical $\pi/2$ pulse length was
$3~\mathrm{\mu s}$. The $^7$Li NMR spectra for narrow lines (FWHM
$\lesssim 100$~kHz) were measured by Fourier transform of half the Hahn
echo signals, while for wider lines, the spectra were measured by
integrating the echo area as a function of the applied magnetic field at
a fixed frequency of rf pulses. Nuclear spin-lattice relaxation rates
were measured by monitoring the recovery of the spin echo height using
the standard saturation-recovery pulse sequence.

\section{\label{scn:results}Results}

\subsection{Magnetic Defect Concentrations\label{Sec:ndefect}}

\begin{figure}[tbp]
\centering\includegraphics[width=3.3in]{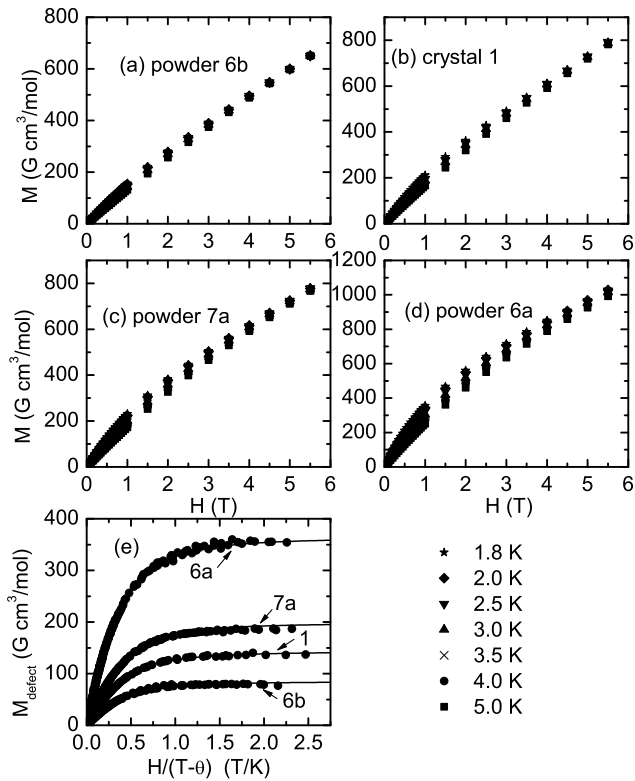}
\caption{(a)--(d) Magnetization $M$ versus applied magnetic field $H$
isotherms at different temperatures $T$ for powder and crystals samples
of LiV$_2$O$_4$.  (e) The magnetic defect contributions
$M_{\mathrm{defect}} = M-\protect\chi_0 H$ to the data in panels (a)--(d)
versus $H/(T-\protect\theta)$.  The 
$\protect \chi_0$ and $\protect\theta$ values are listed in
Table~\protect\ref{tbl:Mfit}.  The solid lines are plots of the second
term in Eq.~(\protect\ref{eqn:M}), $M_{\mathrm{defect}}$, versus
$H/(T-\protect\theta)$ with values of $n_{\mathrm{defect}}$,
$\protect\theta$, and $S$ given in Table~\protect
\ref{tbl:Mfit}.}
\label{fig:mvsh}
\end{figure}

The magnetic defect concentrations of the samples were determined from the
low temperature ($1.8~\mathrm{K} \leq T \leq 5$~K) magnetization $M$
versus applied magnetic field $H$ isotherms.\cite{Kondo1999}
Figures~\ref{fig:mvsh}(a),
\ref{fig:mvsh}(b), \ref{fig:mvsh}(c), and \ref{fig:mvsh}(d) show the
$M(H)$ isotherms at different temperatures for samples 6b, 1, 7a, and 6a,
respectively. The magnetic defect concentration $n_{\mathrm{defect}}$ and
spin value $S$ of the magnetic defects in each sample are determined by
fitting the equation 
\begin{eqnarray} 
M(H,T) &=& \chi_0 H + n_{\mathrm{defect}} N_{\mathrm{A}} g
\mu_{\mathrm{B}} S B_S(x)\notag \\
&\equiv& \chi_0 H + M_{\rm defect}(H,T)
\label{eqn:M}
\end{eqnarray} 
to all the $M(H,T)$ isotherm data for each sample at
$T \leq 5$~K.\cite{Kondo1999} In Eq.~(\ref{eqn:M}),
$n_{\mathrm{defect}}$ is the magnetic defect concentration in
dimensionless mole fraction units, $\chi_0$ is a field- and
temperature-independent contribution to the molar susceptibility at low
temperatures $T \leq 5$ K, $N_{\mathrm{A}}$ is Avogadro's number,
$g$ is the (powder-averaged) spectroscopic splitting factor ($g$-factor)
for the defect spins, $B_S(x)$ is the Brillouin function for spin $S$,
and $x \equiv g \mu_{\mathrm{B}} S H/[k_{\mathrm{B}} (T-\theta)]$ where
$k_{\mathrm{B}}$ is Boltzmann's constant.  We have replaced $T$ in the
usual Brillouin function by $(T-\theta)$ in order to take into account
weak interactions between the magnetic defects.  $\chi_0$, $S$, $\theta$,
and $n_{\mathrm{defect}}$ are free parameters in the fit, whereas the $g$
value is fixed to be equal to 2 during the fit.\cite{Kondo1999}  The best
fit parameters are listed in Table
\ref{tbl:Mfit}.  Figure~\ref{fig:mvsh}(e) shows the magnetic defect
magnetization contributions $M_{\mathrm{defect}}(H,T) = M(H,T) -\chi_0 H$
versus $H/(T-\theta)$ for the four samples.  All the data points in
Figs.~\ref{fig:mvsh}(a), \ref{fig:mvsh}(b), \ref{fig:mvsh}(c), and
\ref{fig:mvsh}(d) fall onto a universal curve in Fig.~\ref{fig:mvsh}(e)
for each sample, respectively, as described by the second term in
Eq.~(\ref{eqn:M}), thus confirming the consistency of the fits.  The
fitted functions $M_{\rm defect}(H,T)$ in Eq.~(\ref{eqn:M}) for the four
samples are plotted versus $H/(T - \theta)$ as the solid curves in
Fig.~\ref{fig:mvsh}(e) and show excellent agreement with the data.

\begin{table}[tbp]
\caption{ Best fit values of the magnetic defect concentration
$n_{\mathrm{defect}}$, the spin value $S$, the intrinsic susceptibility
$\protect\chi_0$, and the effective Weiss temperature $\protect\theta$
for powder samples 6b, 7a, and 6a and crystal sample 1 of
LiV$_2$O$_4$ obtained by fitting Eq.~(\protect\ref{eqn:M}) to the low
temperature (1.8~K $\le T\le 5$~K) magnetization versus field isotherms
in Figs.~\ref{fig:mvsh}(a)--\ref{fig:mvsh}(d) with $0 \leq H \leq
5.5$~T\@.}
\label{tbl:Mfit}
\begin{ruledtabular}
\begin{tabular}{ccccc} 
Sample & $n_{\rm defect}$~(mol\%) & $S$ & $\chi_{0}$ (cm$^3$/mol) &
$\theta $ (K)\\  
\hline 6b & 0.21(1) & 3.6(2) & 0.0104(1) & $-0.75$(14) \\  
7a & 0.49(1) & 3.5(1) & 0.0108(1) & $-0.57$(6) \\  
6a & 0.83(3) & 3.9(1) & 0.0122(2) & $-0.64$(10) \\  
1 & 0.38(1) & 3.3(1) & 0.01186(4) & $-0.43$(6) \\
\end{tabular}
\end{ruledtabular}
\end{table}

Several features of the data in Table~\ref{tbl:Mfit} are important.  The
low-temperature field-independent (up to 5.5~T) susceptibilities $\chi_0$
of all four samples are the same to within about 10\%, even though the
magnetic defect concentrations change by a factor of 4, and are about the
same as in magnetically pure LiV$_2$O$_4$.\cite{Kondo1997,Kondo1999}  This
agreement suggests that the heavy Fermi liquid in magnetically pure
LiV$_2$O$_4$ survives in the presence of the magnetic defects.  Second,
the spins $S$ of the magnetic defects should be considered as average
values, and these values are large, ranging from 3.3 to 3.9.  That the
magnetic defects have large spins is obvious from the data in
Fig.~\ref{fig:mvsh}(e) because the magnetic defect magnetizations are
nearly saturated at relatively low fields of only $\sim 2$~T; spins~1/2
would not saturate even at our maximum field of 5.5~T\@.  It is difficult
to understand how such large spin values could arise from point defects in
the crystal structure.  In that case one might expect the magnetic defect
spins to be much smaller and similar to those of V$^{+4}$ ($S = 1/2$) or
V$^{+3}$ ($S = 1$).  The large spins of the magnetic defects thus
suggest that these spins may be associated with extended objects 
that we call ``magnetic droplets'' in Sec.~\ref{Sec:Droplets} below
instead of being associated with point-like local
magnetic moments as in the usual picture.  Third, the Weiss
temperatures $\theta$ for all the samples are rather small, and indicate
that the average interaction energy between the magnetic defects is also
small and of order 1~K\@.  Finally, from low-temperature magnetization
measurements on many polycrystalline and single crystal samples that we
have carried out in addition to those described here, the magnetic defect
concentrations found do not exceed the largest value listed in
Table~\ref{tbl:Mfit} of 0.83~mol\%.

\subsection{ac Magnetic Susceptibility at 14~MHz}

\begin{figure}[tbp]
\centering 
\includegraphics[width=3in]{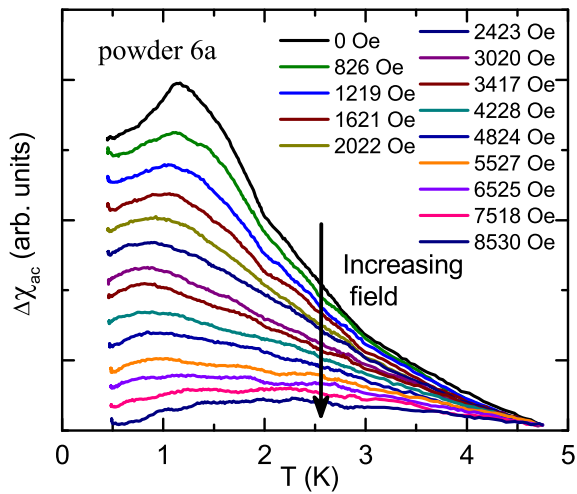}
\caption{(color online.) $\Delta\protect\chi_{\mathrm{ac}}$, the change of
ac magnetic susceptibility $\protect\chi_{\mathrm{ac}}$ at 14~MHz relative
to its value at 4.8~K, versus temperature $T$ for LiV$_{2}$O$_{4}$ powder
sample~6a with $n_{\mathrm{defect}}=0.83$~mol\% at several values of the
magnetic field (indicated in the legend). $\protect\chi_{\mathrm{ac}}$
decreases with increasing magnetic field.}
\label{rpFig2}
\end{figure}

The ac magnetic susceptibility, $\chi_{\mathrm{ac}}=dM/dH$, is an
important parameter directly related to the electronic spin dynamics. It
is very sensitive to collective behavior such as spin freezing and a
transition to the glassy state. Figure \ref{rpFig2} shows
$\Delta\chi_{\mathrm{ac}}\equiv
\chi_{\mathrm{ac}}(T)-\chi_{\mathrm{ac}}(4.8~\mathrm{K})$ versus
temperature 
$T$ at various values of the external magnetic field for powder sample~6a
with $n_{\mathrm{defect}}=0.83$~mol\%.  Each curve corresponds to a
magnetic field listed in the legend and the curves from top to bottom
correspond to increasing magnetic field.  We note that the change of the
magnetic moment amplitude of the measured sample on decreasing the
temperature from 1.1~K to 0.5~K in zero field corresponds to a change in
magnetic moment of only about $5\times 10^{-10}$~G~cm$^3$, which cannot
be resolved by a conventional SQUID magnetometer for the same size ($\sim
0.3$~mm$^3$) sample.

At zero static applied field, there is an obvious peak in $\Delta
\chi_{\mathrm{ac}}$ at about $1.1$~K in Fig.~\ref{rpFig2} that is most
likely indicative of a collective freezing of the magnetic moments. The
field dependence of the magnetic susceptibility is characteristic of a
spin glass system where spin randomness is suppressed by the uniaxial
field and the peak in $\chi_{\mathrm{ac}}$ associated with spin freezing
is suppressed because the magnetic moments are closer to saturation. This
result suggests collective freezing behavior of the magnetic defects in
the LiV$_{2}$O$_{4}$ powder sample in zero field.

\begin{figure}[tbp]
\centering 
\includegraphics[width=3in]{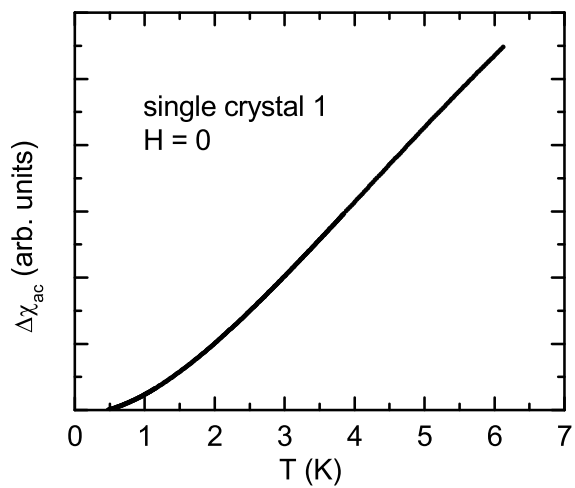}
\caption{$\Delta\protect\chi_{\mathrm{ac}}$, the change of ac magnetic
susceptibility $\protect\chi_{\mathrm{ac}}$ relative to its value at
0.5~K, versus temperature $T$ in single crystal LiV$_{2}$O$_{4}$
(sample~1) in zero applied magnetic field and at a frequency of 14 MHz.}
\label{rpFig3}
\end{figure}

For our sample 1 consisting of a collection of single crystals with
overall $n_{\mathrm{defect}}=0.38$~mol\%, the situation is quite
different.  We cannot measure the spin susceptibility because the
diamagnetic orbital susceptibility $\chi_{\mathrm{ac,skin}}$ arising from
skin depth effects dominates it.  The skin depth $\delta$ can be
calculated from \cite{Corson} 
\begin{equation}
\delta =\frac{504}{(\sigma K_{m}\nu )^{1/2}}~\mathrm{meters},
\label{eqn:skin}
\end{equation}
where $K_{m}$ is relative permeability, $\sigma$ is the
conductivity in $\Omega ^{-1}$m$^{-1}$, and $\nu$ is the applied
frequency in Hz.  Setting $K_{m}=1$, $\sigma =5\times 10^{6}$~$\Omega
^{-1}$m$^{-1}$($\sigma$ value at 1.8~K in Ref.~\onlinecite{Das2007}),
and $\nu =14$~MHz, we obtain $\delta\approx 0.06$~mm, significantly
smaller than the size of a crystal in sample 1.  Thus we expect that
the $\chi_{\mathrm{ac,\,skin}}$ contribution to $\chi _{\mathrm{ac}}$ is
significant and its effect increases with decreasing temperature as the
resistivity decreases monotonically with decreasing
temperature.\cite{Takagi1999,Urano2000}  Figure~\ref{rpFig3} shows the
$\Delta \chi _{\mathrm{ac}}\equiv \chi _{\mathrm{ac}}(T)-\chi
_{\mathrm{ac}}(0.5~\mathrm{K})$ versus temperature $T$ from 0.5 to
6~K\@.  Since the static susceptibility of various samples is nearly
$T$-independent or increases with decreasing $T$ over this $T$ range, the
decrease in $\Delta \chi_{\mathrm{ac}}$ with decreasing $T$ in
Fig.~\ref{rpFig3} indicates that $\chi_{\mathrm{ac,\,skin}}(T)$
dominates the $\chi_{\mathrm{ac}}$ response there.  Furthermore, we see
no evidence for a collective spin freezing for this sample, and we did
not find any field dependence up to an applied field of 10~kOe (not
shown).  Thus, the measurement of the ac susceptibility at 14 MHz for our
single crystals does not yield useful information for understanding
the magnetic response of the magnetic defects in these crystals. 
$\chi_{\rm ac}(H,T)$ measurements at much lower frequencies are
called for.

\subsection{$^7$Li NMR Line Width}

\begin{figure}[tbp]
\centering\includegraphics[width=3in]{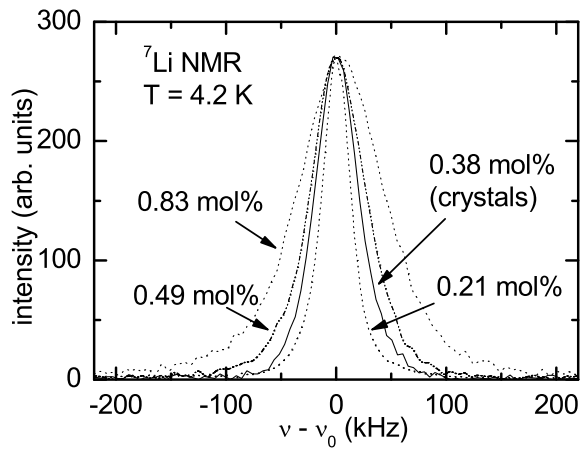}
\caption{The $^7$Li NMR absorption versus rf frequency $\protect\nu$ at
temperature $T=4.2$~K and applied magnetic field $H=1.06$~T in the four
LiV$_2$O$_4$ samples.  The frequency $\protect\nu_0=17.6$~MHz.}
\label{fig:spectrum}
\end{figure}

\begin{figure}[tbp]
\centering\includegraphics[width=3in]{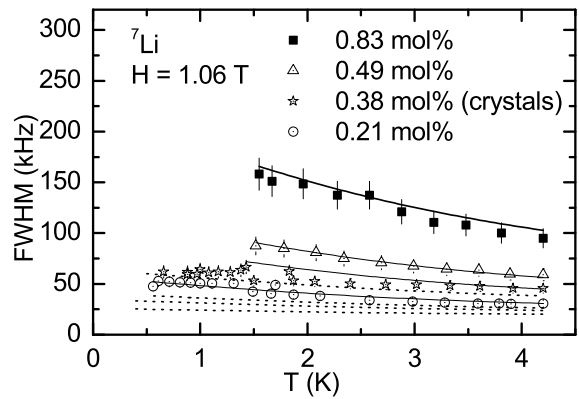}
\caption{Temperature $T$ dependence of full width at half maximum
peak intensity (FWHM) of the $^7$Li NMR spectrum under external magnetic
field $H=1.06$~T in the four LiV$_2$O$_4$ samples.  The symbols are
experimental results.  The dashed lines are plots of
Eq.~(\protect\ref{eqn:FWHMa}) (with $B=1.43$) that takes into account the
contributions due to powder broadening and nuclear dipole-dipole
interactions, but does not take into account local field inhomogeneity
due to the magnetic defects.  The solid lines are fits by
Eq.~(\protect\ref{eqn:FWHMt}), which also takes into account the local
field inhomogeneity.  The fitted solid lines from bottom to top are for
samples with $n_{\mathrm{defect}}=0.21$, 0.38 (crystals), 0.49, and
0.83~mol\%, respectively.}
\label{fig:FWHM}
\end{figure}

The $^7$Li NMR absorption line width is related to the local static
magnetic field distribution. It becomes broader with increasing
concentrations of magnetic defects. Figure~\ref{fig:spectrum} shows the
absorption lines of the four samples at temperature $T=4.2$~K and
$H=1.06$~T\@. Although the $^7$Li nuclei have spin $I=3/2$, both first
and second order nuclear quadrupole broadening due to a structural
distortion can be ruled out since we observe no satellite peaks or
shortening of $\pi/2$ pulse length as compared to the magnetically pure
LiV$_2$O$_4$ sample.\cite{Fukushima,Abragam} The line width is
significantly larger than the intrinsic width for an individual $^7$Li
nuclear spin, indicating an inhomogeneous magnetic broadening of the
line. The intrinsic line width is of the order of $1/T_2 \approx 5$ kHz,
where $T_2$ is the nuclear spin-spin relaxation time and is almost
independent of the defect concentration and temperature below 4.2 K\@.
Figure \ref{fig:FWHM} displays the temperature dependences of the full
width at half maximum peak intensity (FWHM) of the spectra for the four
samples.

The broadening of the $^7$Li NMR line has three contributions. The first
contribution comes from the nuclear $^7$Li-$^{51}$V and $^7$Li-$^7$Li
dipolar interactions.  This contribution can be estimated using the Van
Vleck second moment $\langle \Delta \omega^2 \rangle$.\cite{Vleck1948}  A
second broadening comes from the macroscopic field inhomogeneity due to a
distribution of the demagnetization factors and a distribution of magnetic
fields due to neighboring powder grains.  This contribution is
proportional to the magnetization of the sample and the resulting root
mean square deviation of $^7$Li NMR resonance frequencies can be written
as $B M \rho_N \gamma_{\mathrm{Li}}/2\pi$, where $M$ is the molar
susceptibility, $\rho_N$ is the density of LiV$_2$O$_4$ formula units in
the sample, $\gamma_{\mathrm{Li}}$ the gyromagnetic ratio of $^7$Li
nuclei, and $B$ a dimensionless factor.  $B$ is estimated to be 1.43 for a
close packed powder sample with ellipsoidal shapes.\cite{Drain1962}  A
third broadening contribution comes from inhomogeneity due to the
presence of magnetic defects within the sample.  An estimate for this
contribution is not possible without a model of the nature of the defects
and the types of interactions between the defects and nearby $^7$Li
nuclear spins.  However, the presence of this contribution can be inferred
by comparing the experimental FWHM values and the values expected when
including only the first two contributions, as follows.

The FWHM resulting from the first two contributions can be calculated
within a Gaussian approximation by 
\begin{equation}
\mathrm{FWHM}_{\mathrm{a}}=2.35\sqrt{\langle \Delta \omega^2
\rangle/(2\pi)^2+(B M \rho_N \gamma_{\mathrm{Li}}/2\pi)^2} 
\label{eqn:FWHMa}
\end{equation} with $B=1.43$, and $\langle \Delta \omega^2
\rangle^{1/2}/2\pi=2.7$~kHz.\cite{Onoda1997} $M$ is calculated from
Eq.\ (\ref{eqn:M}) using the parameter values listed in Table
\ref{tbl:Mfit}. The FWHM$_{\mathrm{a}}$ calculated from
Eq.~(\ref{eqn:FWHMa}) is plotted as the dashed lines in
Fig.~\ref{fig:FWHM}. It is clear that Eq.~(\ref{eqn:FWHMa}) cannot
account for the observed broadening of the lines, so a local magnetic
field inhomogeneity due to the presence of the magnetic defects must be
present in the samples.  We will return to this issue in
Sec.~\ref{sec:linewidth}.

\subsection{Nuclear Spin-Lattice Relaxation Rates}

\begin{figure}[tbp]
\centering
\includegraphics[width=3.2in]{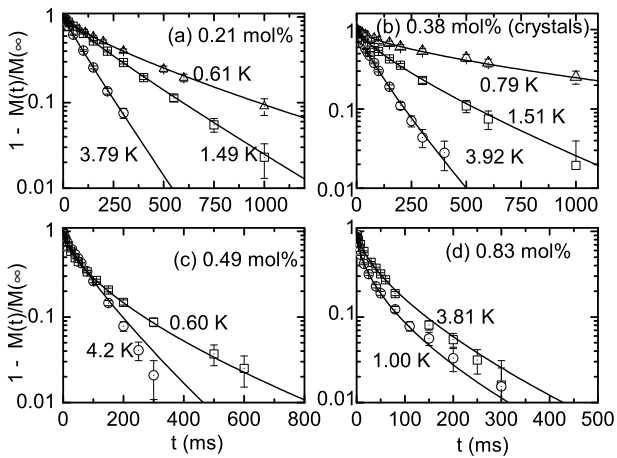}
\caption{Recovery of $^{7}$Li nuclear magnetization $M(t)$ after time
delay $t$ following a sequence of saturation pulses.  Note that $M$
here is different from the electronic spin magnetization in
Fig.~\protect\ref{fig:mvsh}.  The data points were obtained in applied
magnetic field $H=1.06$~T at the indicated temperatures and with rf
frequency $\protect\nu =17.6$~MHz for LiV$_{2}$O$_{4}$ samples with (a)
0.21~mol\%, (b) 0.38~mol\% (single crystals), (c) 0.49~mol\%, and (d)
0.83~mol\% magnetic defects.  The solid curves are fits to the data by
Eq.~(\protect\ref{eqn:strfit}).}
\label{fig:rlxall}
\end{figure}

\begin{figure}[tbp]
\centering
\includegraphics[width=2.5in]{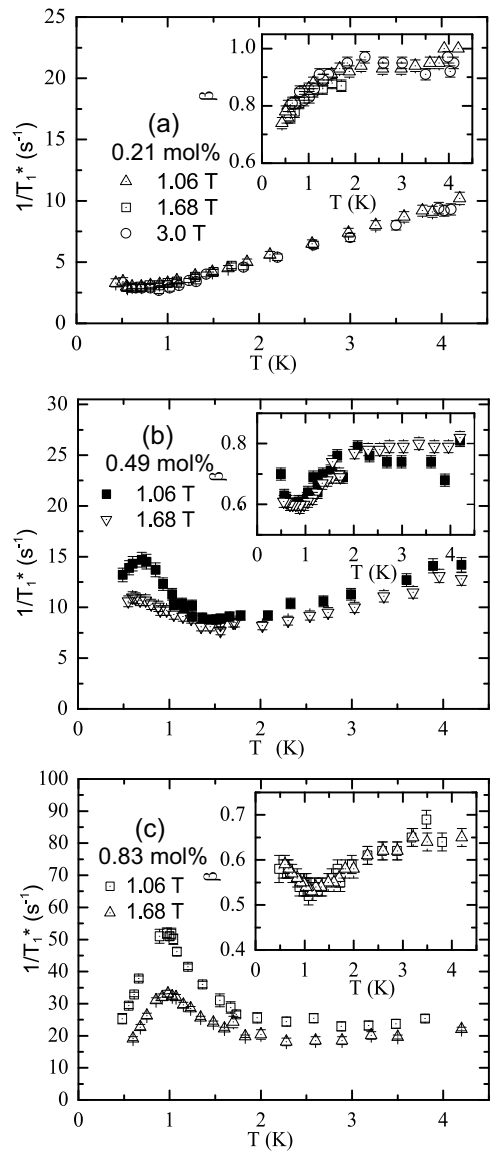}
\caption{$1/T_{1}^{\ast}$ and $\protect\beta$ versus temperature $T$
of LiV$_{2}$O$_{4}$ obtained by fitting data as in
Figs.~\protect\ref{fig:rlxall}(a), (c), and (d) by
Eq.~(\protect\ref{eqn:strfit}), of (a) powder sample~6b with
$n_{\mathrm{defect}}=0.21$~mol\% at external magnetic fields $H= 1.06$,
1.68, and 3.0~T, (b) powder sample~7a with
$n_{\mathrm{defect}}=0.49$~mol\% at $H= 1.06$, 1.68~T, and (c) powder
sample~6a with $n_{\mathrm{defect}}=0.83$~mol\% at $H= 1.06$ and
1.68~T\@.}
\label{fig:IT1beta}
\end{figure}

\begin{figure}[tbp]
\centering
\includegraphics[width=2.5in]{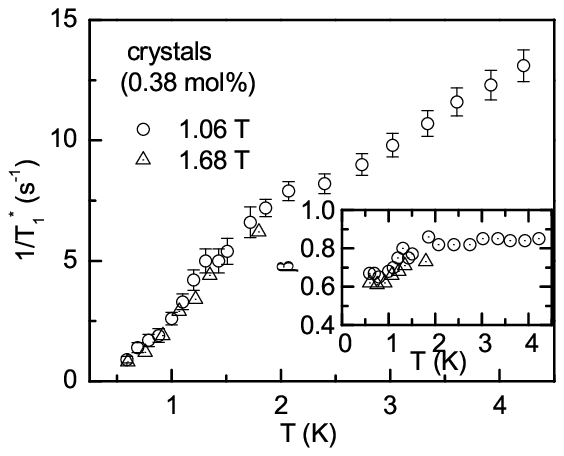}
\caption{$1/T_{1}^{\ast}$ and $\protect\beta$ versus temperature $T$ of
the LiV$_{2}$O$_{4}$ crystal sample 1 with
$n_{\mathrm{defect}}=0.38$~mol\% in external magnetic fields $H= 1.06$
and 1.68~T, obtained by fitting data as in
Fig.~\protect\ref{fig:rlxall}(b) by Eq.~(\protect\ref{eqn:strfit}).}
\label{fig:IT1beta-xstal}
\end{figure}

The longitudinal $^7$Li nuclear spin relaxation versus time $M(t)$
exhibits an increasingly nonexponential behavior with increasing
concentration of magnetic defects or decreasing temperature. 
Figure~\ref{fig:rlxall} shows the recoveries of $M(t)$ following a
saturation sequence for the four samples at different temperatures.  The
recovery data can be described by a stretched exponential function 
\begin{equation}
1-\frac{M(t)}{M(\infty)}=\mathrm{exp}[-(t/T_{1}^{\ast})^{\beta}].
\label{eqn:strfit}
\end{equation} 
The solid curves in Fig.~\ref{fig:rlxall} are best fits to
the data by Eq.~(\ref{eqn:strfit}). The best fit values of
$1/T_{1}^{\ast}(T)$ and $\beta(T)$ are shown in Figs.~\ref{fig:IT1beta}
and \ref{fig:IT1beta-xstal} for powder and single crystal samples,
respectively.

The temperature dependence of $1/T_1^*$ is quite different in the powder
and single crystal samples.  A peak is observed in $1/T_1^*(T)$ for the
powder samples 6a~($n_{\mathrm{defect}}=0.83$~mol\%,
$T_{\mathrm{peak}}\approx1.0$~K) and
7a ($n_{\mathrm{defect}}=0.49$~mol\%,
$T_{\mathrm{peak}}\approx0.6$--0.7~K). In the powder sample~6b with the
smallest magnetic defect concentration
($n_{\mathrm{defect}}=0.21$~mol\%), $1/T_1^*$ starts to increase at the
lowest experimental temperatures and might exhibit a peak with further
decreasing temperature.  The peak positions in sample~6a for
$H=1.06$ and 1.68~T are almost the same as the peak position in
$\chi_{\mathrm{ac}}(T)$ for this sample at $H=0$ in Fig.~\ref{rpFig2}.
We conclude that the peaks in $1/T_1^*$ originate from the spin freezing
of the magnetic defects.  In the crystal sample, $1/T_1^*(T)$ in
Fig.~\ref{fig:IT1beta-xstal} decreases monotonically with decreasing
temperature with a $1/T_1^*$ value at 0.5~K much smaller than in the
powder samples, and there is no sign of spin freezing.

\begin{figure}[tbp]
\centering
\includegraphics[width=3in]{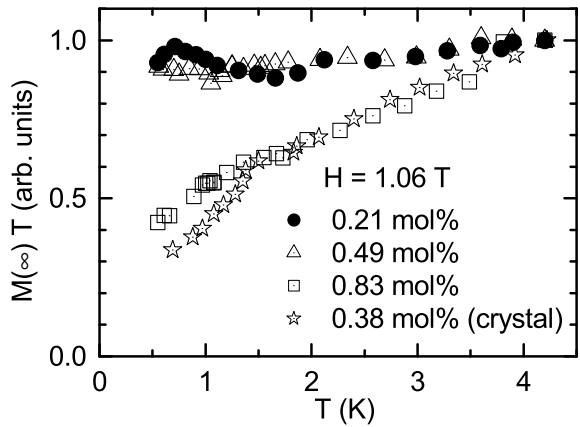}
\caption{The fully recovered echo intensity $M(\infty)$, which is the
total equilibrium nuclear magnetization, times temperature $T$ versus $T$
in the nuclear spin-lattice relaxation rate measurements of the four
LiV$_2$O$_4$ samples in an applied field $H=1.06$~T\@.}
\label{fig:intensity}
\end{figure}

Before ending this subsection, we comment about the effect of
inhomogeneous broadening on the relaxation measurements.  Because of the
increasing inhomogeneous broadening with decreasing temperature, some of
the $^7$Li nuclei may be shifted out of the NMR spectrometer response
window ($\Delta f\sim 200$~kHz) and excluded from the relaxation
measurements.  The number of observed $^7$Li nuclei can be estimated from
the product of fully recovered echo height $M(\infty)$ and the
temperature, which is proportional to the nuclear Curie constant $C$ in
the Curie law for $M(\infty)=C/T$.  These data are shown for $H=1.06$~T
versus temperature $T$ in Fig.~\ref{fig:intensity}.  For powder samples~6b
($n_{\mathrm{defect}}=0.21$~mol\%) and 7a
($n_{\mathrm{defect}}=0.49$~mol\%), the decrease of $M(\infty)T$ is less
than 10\% when the temperature decreases from 4.2~K to the lowest
temperature ($\approx0.5$~K).  In contrast, for sample~6a
($n_{\mathrm{defect}}=0.83$~mol\%), $M(\infty)T$ starts to decrease below
$T\approx3.5$~K and at the lowest temperature ($T\approx 0.5$~K),
$M(\infty)T$ is about 50\% of that at 4.2~K\@.  As we will show below, the
nuclei at the wings of the spectrum have an average relaxation rate
larger than those at the center of the spectrum.  Exclusion of those
nuclei in sample~6a can thus result in a smaller measured relaxation rate
in that sample.

In the single crystals, the normalized signal intensity $M(\infty)T$ also
decreases with decreasing temperature.  Since the line width in the
crystals is less than in powder sample~7a (see Fig.~\ref{fig:FWHM}),
where no significant signal loss is observed, we attribute the signal
loss to the effect of rf field skin depth.  Here, only the $^7$Li nuclear
spins within the skin depth contribute to the NMR signal.  Setting $K_m
=1$, $\sigma=5\times 10^6~\Omega^{-1}$m$^{-1}$ (the value of $\sigma$ at
1.8~K in Ref.~\onlinecite{Das2007}), and $\nu=17.6$~MHz,
Eq.~(\ref{eqn:skin}) gives $\delta =0.054$~mm, which is less than the
typical size (0.2~mm) of the crystals.  However, there is an unexplained
kink in the data for the crystals at $T\approx 1.4$~K in both
Figs.~\ref{fig:IT1beta-xstal} and \ref{fig:intensity}.

\subsection{Relaxation at Different Positions in the Spectra}

The observation of a stretched exponential relaxation behavior indicates
the presence of a distribution of nuclear spin-lattice relaxation rates
$1/T_1$. In order to study the origin of the $1/T_1$ distribution, we
performed the following ``hole burning" experiment. This experiment
extends our previous hole burning experiment briefly described in
Ref.~\onlinecite{Johnston2005}. We also studied the relaxation behavior
at different positions of the NMR absorption line.

\begin{figure}[tbp]
\centering\includegraphics[width=3.0in]{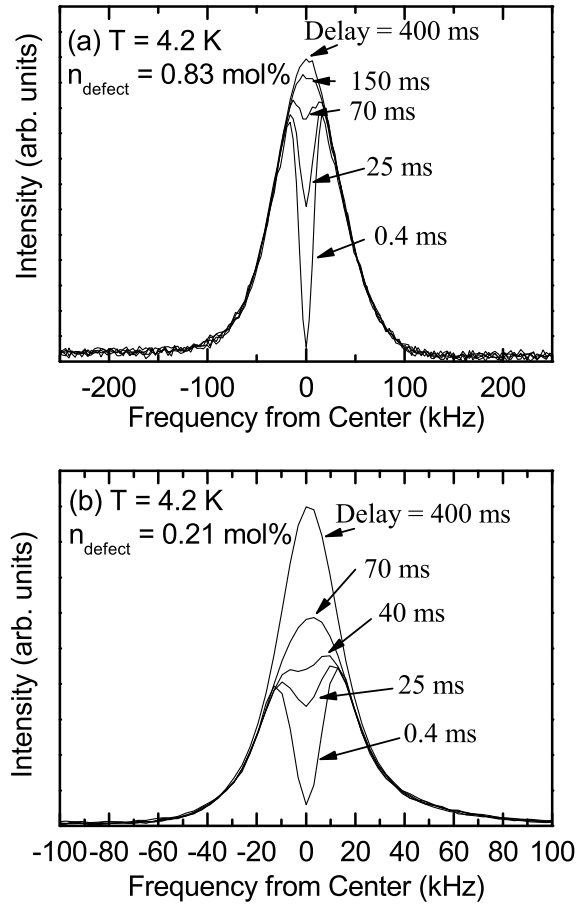}
\caption{Recovery at 4.2~K of a ``hole'' in the absorption spectrum of
LiV$_{2}$O$_{4}$ produced by a weak $\protect\pi/2$ pulse with pulse
length of 56~$\mathrm{\protect\mu s}$ at delay = 0 in sample (a)~6a
($n_{\mathrm{defect}}=0.83$~mol\%) and (b)~6b
($n_{\mathrm{defect}}=0.21$~mol\%).  The applied magnetic field $H$ is
1.06~T and the center frequency is 17.6~MHz.  The delay times after which
the spectra were measured by two strong rf pulses are given in the
figures.  Note the different abscissa scales in (a) and (b).}
\label{fig:holerecovery}
\end{figure}

Figures~\ref{fig:holerecovery}(a) and~(b) display the recovery of a
``hole'' in the echo spectrum in applied magnetic field $H=1.06$~T,
obtained from Fourier transform of half the Hahn echo signal generated by
two strong rf pulses following a weak $\pi/2$ pulse in samples~6a
($n_{\mathrm{defect}}=0.83$~mol\%) and 6b
($n_{\mathrm{defect}}=0.21$~mol\%), respectively.  The weak $\pi/2$ pulse
has a width of 56~$\mathrm{\mu s}$ and most of its power is distributed
within a narrow frequency window of width $\approx40$~kHz.  Such a weak
$\pi/2$ pulse only saturates the central part of the spectrum.  It is
clear that the hole recovery process does not affect the rest of the line
and thus spectral diffusion does not occur in our time scale.  That is,
nuclei with different Larmor frequencies are not coupled to each other
over the NMR relaxation time scale of $T_1\sim 100$~ms. 

\begin{figure}[tbp]
\centering
\includegraphics[width=3in]{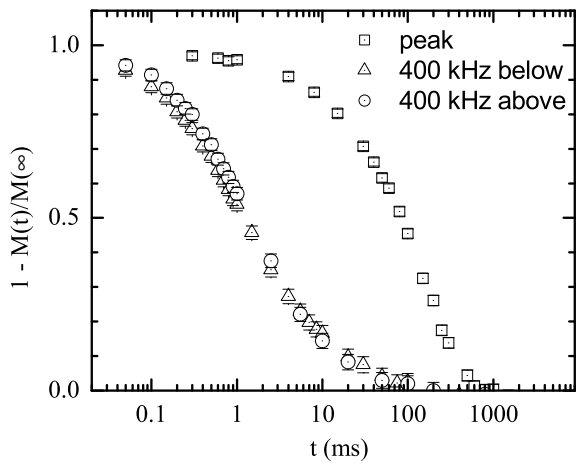}
\caption{ Recovery at 4.2~K of $^7$Li longitudinal nuclear magnetization
$M(t)$ following a saturation sequence at time $t=0$ measured at different
positions of the spectrum in LiV$_{2}$O$_{4}$ powder sample~7a
($n_{\mathrm{defect}}=0.49$~mol\%) under external magnetic field $H =
1.68$~T\@.  The recovery curves,  which are nonexponential, were measured
with rf pulse frequency ($\boxdot$) equal to the peak of the spectrum
(27.8~MHz) and at $T=1.56$~K, ($\triangle$) 400 kHz lower than the peak
and at $T=1.53$~K, and ({\protect\large $\circ$}{\protect\normalsize)
400~kHz higher than the peak and at $T=1.77$~K, respectively.}}
\label{fig:rlxSD7a}
\end{figure}

Lack of spectral diffusion as observed above allows us to investigate the
nuclear spin-lattice relaxation at different positions of the spectrum.
Due to the strong $^7$Li NMR signal at low temperatures, we were able to
study the relaxation of $^7$Li far out on the wings of the spectrum
although the signal intensity is much weaker than at the peak.
Figure~\ref{fig:rlxSD7a} displays the nuclear spin-lattice relaxation
curves of powder sample~7a ($n_{\mathrm{defect}}=0.49$~mol\%) in $H =
1.68$~T with the rf pulse frequency equal to, 400~kHz higher than, or
400~kHz lower than, the peak frequency of the line.  All three recovery
curves are nonexponential.  It is clear from Fig.~\ref{fig:rlxSD7a} that
the nuclei close to the peak of the line have an average relaxation rate
lower than those away from the peak.  As will be discussed below, the
behavior in Fig.~\ref{fig:rlxSD7a} is consistent with an inhomogeneous
local magnetic field induced by the magnetic defects.  It is noted that
the temperatures at which the three relaxation curves were taken are
slightly different.  However, such small temperature differences should be
negligible compared to the large difference of relaxation rates between
these three curves. 

\section{\label{scn:models}Analysis}

\subsection{Introduction}

A model for the microscopic nature of the magnetic defects has to be
assumed in order to analyze the NMR results.  We will examine two
related possibilities.  First, in Sec.~\ref{sec:geometry} we treat the
magnetic defects as traditional identical localized magnetic
moments.  The distribution of $^7$Li nuclear spin-lattice relaxation
rates and the inhomogeneous broadening of the $^7$Li absorption spectrum
are then entirely caused by the local field inhomogeneity, which arises
from a distribution of positions of the
$^{7}$Li nuclei relative to the magnetic defects.  In a refined version
of this model in Sec.~\ref{Sec:Droplets}, the magnetic defects are
assumed to actually be ``magnetic droplets'' that have a distribution of
sizes.  In this refined approach, the distribution of the dynamic
properties of the droplets also needs to be considered.

We will assume at the outset that the heavy Fermi liquid present in
magnetically pure LiV$_{2}$O$_{4}$ is not affected by the presence of
the magnetic defects.  The measured $^{7}$Li relaxation versus time
following saturation then arises from two mechanisms.  The first is a
single-exponential relaxation that is the same as in magnetically pure 
LiV$_{2}$O$_{4}$ and that comes from the contact interaction of
$^{7}$Li nuclear spins with the conduction electron spins in the heavy
Fermi liquid.  The second mechanism is the hyperfine interaction of the
nuclei with the magnetic defects.  The first (homogeneous) mechanism
gives a relaxation rate described by the Korringa law with
$1/T_{1}\propto T$.\cite{Slichter}  The main goal of the present modeling
is to then  determine the (second) contribution of the magnetic defects
to the time-dependent nuclear relaxation and to subsequently  interpret
what that contribution means.  The separation of these two contributions
to the magnetic properties, at least above 1.8~K, is supported by previous
magnetization measurements\cite{Kondo1997, Kondo1999} as well as by those
in Sec.~\ref{Sec:ndefect} above.  The magnetization as expressed in
Eq.~(\ref{eqn:M}) contains a contribution
$\chi _{0}H$ almost independent of the magnetic defect concentration (see
Table~\ref{tbl:Mfit}).  This contribution is most likely due to the same
heavy Fermi liquid that is present in magnetically pure LiV$_{2}$O$_{4}$
at low temperatures.  

We will see that this separation of the magnetic properties into a heavy
Fermi liquid part and a magnetic defect part can consistently explain
our $^7$Li NMR measurements on our powder samples of
LiV$_{2}$O$_{4}$.  However, as we will show in Fig.~\ref{fig:ptxtal}(a)
below, our NMR longitudinal magnetization recovery data below $\sim
1.3$~K for our sample of \emph{single crystals} indicate that the Fermi
liquid is modified by the presence of magnetic defects and/or that our
model for the magnetic defects is no longer accurate below that
temperature in our single crystals.  This differentiation between the
bulk crystal and powder properties indicates that there are differences
between the natures of the magnetic defects and/or their interactions in
single crystals as compared to powders, which in turn are likely
associated in some way with the quite different preparation conditions of
the two types of samples.

\subsection{Geometric Inhomogeneity\label{sec:geometry}}

\subsubsection{\label{sec:linewidth}$^7$Li NMR Line Width}

First we will analyze the $^7$Li NMR line width by considering the 
distribution of distances between nuclear spins and point-like magnetic
defects within the spinel structure.  Dilute paramagnetic centers give
rise to a broadening of the NMR spectrum through inhomogeneous dipolar
and RKKY interactions and in the limit of great dilution the line shape
approaches a Lorentzian with full width at half maximum intensity 
(FWHM$_{\mathrm{b}}$) given by
\cite{Walstedt1974} 
\begin{eqnarray}
\mathrm{FWHM_{b}} &=&A n_{\mathrm{defect}}\frac{8\pi \rho
_{N}}{9\sqrt{3}} g\mu _{\mathrm{B}}\gamma
_{\mathrm{Li}}\langle S_{z}\rangle \notag \\
&=&4.5An_{\mathrm{defect}}SB_{S}(x)~\mathrm{MHz},  
\label{eqn:FWHMb}
\end{eqnarray}
where $\rho _{N}=1.44\times 10^{22}$~cm$^{-3}$ is the number 
density of LiV$_{2}$O$_{4}$ formula units, $A = 1$ for purely dipolar
interactions and $A>1$ if the RKKY interaction is also important,
$\langle S_{z}\rangle$ is the thermal average value of magnetic defect
spin polarization along the direction of the applied magnetic field and
is equal to $SB_{S}(x)$ with $x=g\mu
_{\mathrm{B}}SH/[k_{\mathrm{B}}(T-\theta)]$ [see Eq.~(\ref{eqn:M})].  The
line shape due to the dilute magnetic defects is
Lorentzian\cite{Walstedt1974} while the line shape due to the two
contributions in Eq.~(\ref{eqn:FWHMa}) is
Gaussian.\cite{Vleck1948,Drain1962}  In order to obtain the final FWHM
value, we convolute a Gaussian distribution with FWHM = 1 with a
Lorentzian distribution that has FWMH = $x$ and the same mean value as
the Gaussian distribution.  We find that the FWHM of the convoluted
distribution can be approximated by $(1+x^{8/5})^{5/8}$ to within 10\%
for all values of $x$. We estimate the total FWHM by combining
Eqs.~(\ref{eqn:FWHMa}) and (\ref{eqn:FWHMb}) according to 
\begin{eqnarray}
\mathrm{FWHM}&=&(\mathrm{FWHM_{a}}^{\frac{8}{5}} +
\mathrm{FWHM_{b}}^{\frac{8}{5}}) ^{\frac{5}{8}} \notag  
\label{eqn:FWHMt} \\ 
&=&\left\{
\mathrm{FWHM_{a}}^{\frac{8}{5}}+[4.5An_{\mathrm{defect}}SB_{S}(x)~
\mathrm{MHz}]^{\frac{8}{5}}\right\}^{\frac{5}{8}}.  \notag \\
\end{eqnarray}
Using the values of $n_{\mathrm{defect}}$, $S$ and $\theta $
in Table~\ref{tbl:Mfit} and the results for
FWHM$_{\mathrm{a}}$ in Fig.~\ref{fig:FWHM}, the FWHM 
data in Fig.~\ref{fig:FWHM} for all four samples were simultaneously
fitted by Eq.~(\ref{eqn:FWHMt}), except for the single crystal data below
1.3~K,  where the nuclear spin-lattice relaxation rates in
Fig.~\ref{fig:IT1beta-xstal} indicate a possible screening of the
magnetic defects.   The only fitting parameter was $A$, and the best fit
value was $A = 1.4$.  The best fit to the data is displayed as the
set of solid curves in Fig.~\ref{fig:FWHM}.  The high quality of the fit
shows that the local field inhomogeneity at the $^7$Li nuclear sites
arising from the distribution of distances between the $^7$Li nuclei and
the magnetic defects is an essential contributor to the $^7$Li NMR line
width.

\subsubsection{$^7$Li Nuclear Spin-Lattice Relaxation\label{Sec:NSLR}}

In the present approach we treat the
magnetic defects as identical localized magnetic
moments.  The distribution of $^{7}$Li nuclear spin-lattice
relaxation rates $1/T_{1}$ then arises from a distribution of 
fluctuating local magnetic fields at the nuclear sites due to a
distribution in the positions of the nuclei relative to the magnetic
defects.  Since the relative positions of the $^{7}$Li nuclei with
respect to the magnetic defects are fixed, the shape of the $^7$Li
$1/T_{1}$ probability distribution due to the defects should be
temperature independent.  This would give rise to a temperature
independent $\beta$ value\cite{Johnston2005, Johnston2006} in the
stretched exponential function in Eq.~(\ref{eqn:strfit}) if there were no
additional contributions to the
$^{7}$Li nuclear spin-lattice relaxation.

The observed temperature dependences of the stretching exponent $\beta$ in
the insets of Figs.~\ref{fig:IT1beta} and \ref{fig:IT1beta-xstal} are
explained in this model by the additional Korringa contribution to
$1/T_1$ that is proportional to the temperature.  Since the nuclear
spin-lattice relaxation rate due to itinerant conduction electrons is
assumed to be homogeneous across the sample since it results from the
contact interaction between the nuclear and conduction electron spins, the
nuclear spin recovery due to the conduction electrons alone should be a
single exponential.  As just discussed, the recovery due to the magnetic
defects alone should be a stretched exponential function with a
temperature-independent
$\beta$. The observed temperature dependent $\beta$ arises in our model
from different temperature dependences of the Korringa and magnetic defect
contributions to the nuclear spin-lattice relaxation.  Different
temperature dependences result in different weights of these two
contributions at different temperatures and accordingly different $\beta$
values are seen at different temperatures when the total recovery is
fitted by a stretched exponential function Eq.~(\ref{eqn:strfit}). 
Similarly, it is not appropriate to analyze the $1/T_1^*(T)$ data in
Figs.~\ref{fig:IT1beta} and~\ref{fig:IT1beta-xstal} in terms of a sum of 
contributions from the heavy Fermi liquid and from the local magnetic
defects, because their respective contributions to $1/T_1^*(T)$ cannot be
deconvoluted.  

To determine the magnetic defect contribution to the $^7$Li nuclear spin
dynamics, we first extract from the observed $^7$Li nuclear
spin-lattice relaxation versus time $M(t)$ data what the contribution of
the magnetic defects is, and \emph{then} derive parameters describing the 
relaxation by the magnetic defects.  To accomplish the former goal, we 
write $1 - M(t)/M(\infty) = p(t)
\exp(-t/T_{1\rm{K}})$, where $p(t)$ is the contribution to the
$^7$Li nuclear spin relaxation from the magnetic defects and
$\exp(-t/T_{1\rm{K}})$ is the Korringa contribution from the heavy Fermi
liquid, where we assume a concentration independent Korringa relaxation 
rate $1/T_{1\mathrm{K}}=(2.2~\mathrm{s}^{-1}\mathrm{K}^{-1})T$
and the coefficient of $T$ is taken to be the value in a magnetically pure
sample.\cite{Johnston2005}  Then one obtains  
\begin{equation}
p(t)=\left[1-\frac{M(t)}{M(\infty)}\right]
\exp\left(\frac{t}{T_{\mathrm{1K}}}\right).
\label{eqn:pt}
\end{equation} 
Thus $p(t)$ is determined by multiplying the experimentally observed $1-
M(t)/M(\infty)$ by $\exp(t/T_{\mathrm{1K}})$.

\begin{figure}[tbp]
\centering
\includegraphics[width=3in]{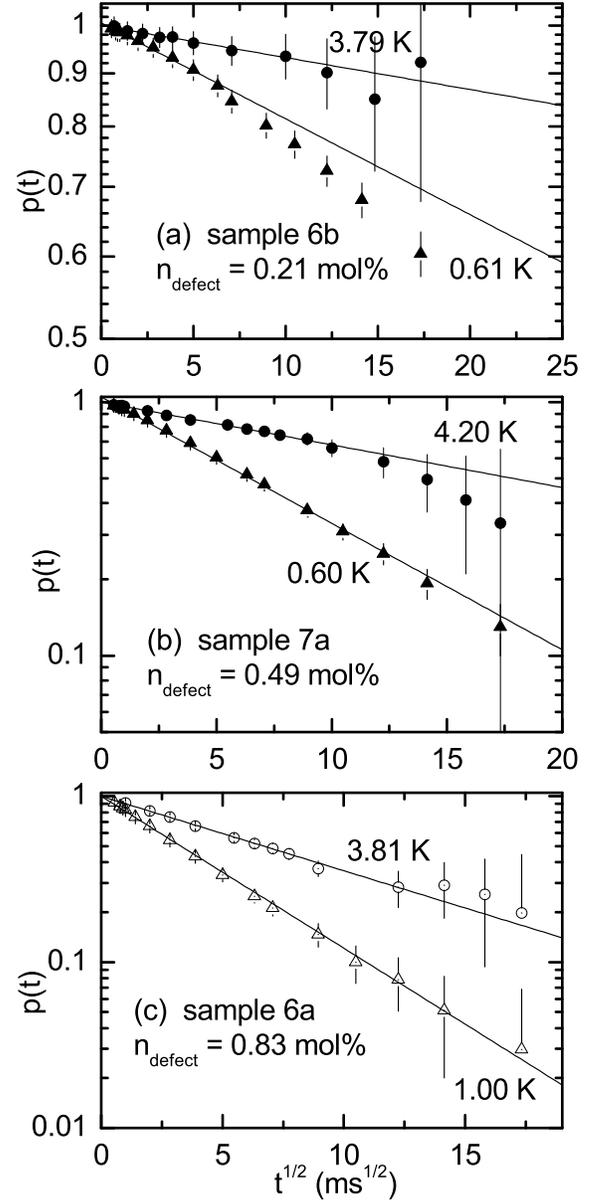}
\caption{Semilog plot of $p(t)$ in Eq.~(\protect\ref{eqn:pt}) versus the
square root of the delay time $t^{1/2}$ for the LiV$_{2}$O$_{4}$ powder
samples (a)~6b with $n_{\mathrm{defect}}=0.21$~mol\%, (b)~7a with
$n_{\mathrm{defect}}=0.49$~mol\%, and (c)~6a with
$n_{\mathrm{defect}}=0.83$~mol\% at applied magnetic field $H=1.06$~T and
different temperatures.  The straight lines are best fits of the data by
Eq.~(\protect\ref{eqn:rootexp}), with parameters $1/T_{\mathrm{1d}}^*$
given in Fig.~\protect\ref{fig:T1d}.}
\label{fig:ptall}
\end{figure}

\begin{figure}[tbp]
\centering
\includegraphics[width=3in]{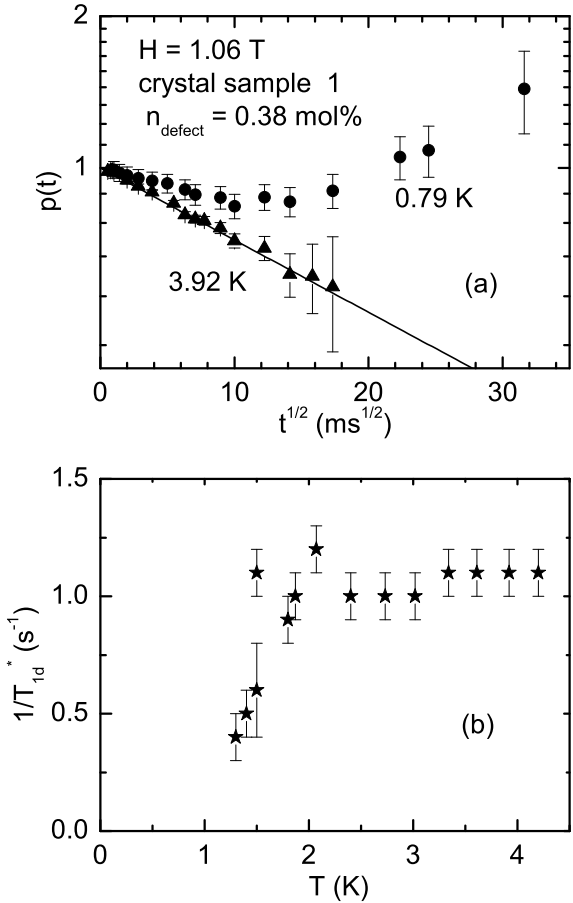}
\caption{(a) Semilog plot of the relaxation function $p(t)$ in
Eq.~(\protect\ref{eqn:pt}) versus the square root of the delay time
$t^{1/2}$ for the LiV$_{2}$O$_{4}$ crystal sample~1 with
$n_{\mathrm{defect}}=0.38$~mol\% at applied magnetic field $H=1.06$~T and
at two different temperatures. The upturn in the 0.79~K data at large
times is unphysical (see text).  The straight line is a
best fit of the 3.92~K data by Eq.~(\protect\ref{eqn:rootexp}). (b)
$1/T_{\mathrm{1d}}^*$ in Eq.~(\protect\ref{eqn:rootexp}) versus
temperature $T$ in $H=1.06$~T and above 1.3~K, where the upturn seen for
$T=0.79$~K in (a) is absent.}
\label{fig:ptxtal}
\end{figure}

We find that the magnetic defect contribution $p(t)$ to the $^7$Li
nuclear spin-lattice relaxation usually follows a stretched exponential
time dependence in our temperature range $0.5 \leq T \leq 4.2$~K with a
temperature- and magnetic defect concentration-independent stretching
exponent $\beta$, as anticipated above, where we find that $\beta$ has the
specific value  $\beta = 1/2$.  Thus we obtain
\begin{equation} 
p(t)=\exp[-(t/T_{\mathrm{1d}}^{\ast})^{1/2}] ,
\label{eqn:rootexp}
\end{equation} 
where the new parameter $T^*_{\mathrm{1d}}$ takes the place of $T_1^*$ in
Eq.~(\ref{eqn:strfit}).  Figures~\ref{fig:ptall} and
\ref{fig:ptxtal}(a) show plots of the logarithm of $p(t)$ versus
$t^{1/2}$ in external magnetic field $H=1.06$~T and at different
temperatures for powder and single crystal samples, respectively.  In
powder samples~7a ($n_{\mathrm{defect}}=0.49$ mol\%) and~6a
($n_{\mathrm{defect}}=0.83$ mol\%), $p(t)$ can be fitted very well by
Eq.~(\ref{eqn:rootexp}) at all temperatures, as shown by the linear fits
in Figs.~\ref{fig:ptall}(b) and~(c), respectively.  In powder sample~6b
with a smaller $n_{\mathrm{defect}}=0.21$~mol\% in
Fig.~\ref{fig:ptall}(a), $p(t)$ follows root exponential behavior for
all times $t$ at the higher temperatures but only at short times at the
low temperature of 0.61~K\@.  We infer in
Sec.~\ref{Sec:SpinDiffusion} below that the deviation at longer times is
due to the effect of spin diffusion.  In the crystals, $p(t)$ in
Fig.~\ref{fig:ptxtal}(a) follows root exponential decay above 1.3~K but
at lower temperature $p(t)$ instead shows an unphysical increase at later
times.  This unphysical behavior suggests that Eq.~(\ref{eqn:pt})
overestimates the conduction electron contribution to the nuclear
spin-lattice relaxation at temperatures below 1.3~K, the separability
of the relaxation due to the magnetic defects from that due to the
conduction electrons is no longer appropriate in that temperature
regime, or the conduction electrons no longer form a Fermi liquid. An
additional possible reason for the unphysical behavior is given in
Sec.~\ref{Sec:Droplets}.  Resolving this issue is an important topic for
future research.

\begin{figure}[tbp]
\centering
\includegraphics[width=3in]{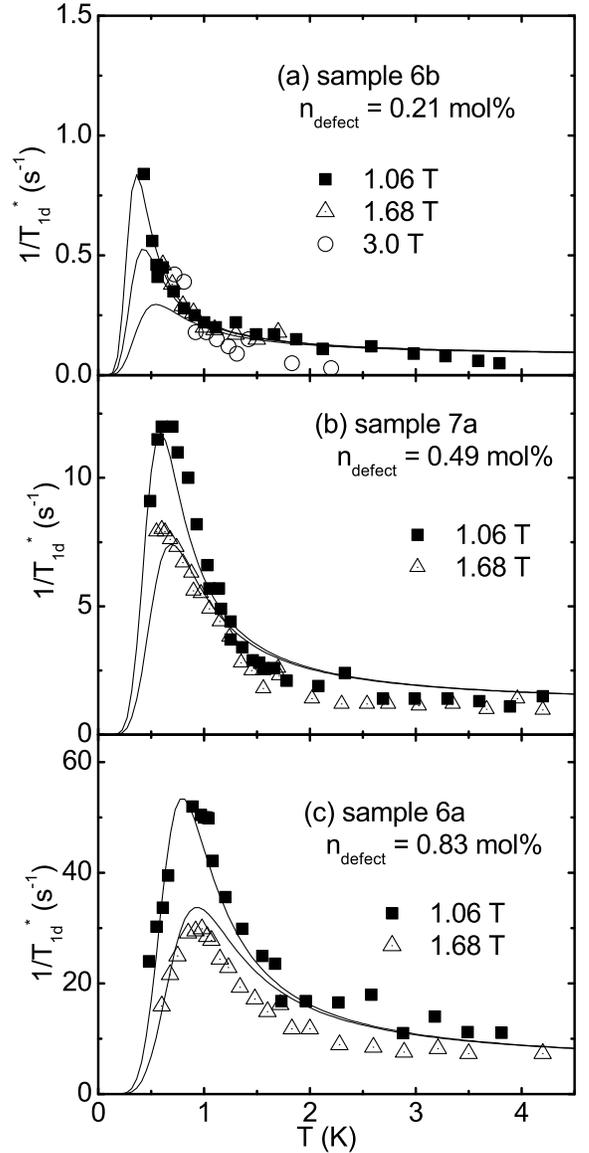}
\caption{$1/T_{\mathrm{1d}}^{\ast}$ versus temperature $T$ at applied
magnetic fields $H=1.06$, 1.68, and 3.0~T for the LiV$_{2}$O$_{4}$ powder
samples (a)~6b with $n_{\mathrm{defect}}=0.21$~mol\%, (b)~7a with
$n_{\mathrm{defect}}=0.49$~mol\%, and (c)~6a with
$n_{\mathrm{defect}}=0.83$~mol\%.  The solid curves are fits to the data
by Eqs.~(\protect\ref{eqn:acttau}) and (\protect\ref{eqn:Lorentz2}).}
\label{fig:T1d}
\end{figure}

We extract $1/T_{\mathrm{1d}}^{\ast}$ versus temperature from the slopes
of the fitted lines of $\log[p(t)]$ versus $t^{1/2}$ in
Figs.~\ref{fig:ptall} and \ref{fig:ptxtal}(a) according to
Eq.~(\ref{eqn:rootexp}).  The results are displayed in
Figs.~\ref{fig:ptxtal}(b) and
\ref{fig:T1d} for the single crystal and powder samples, respectively.
The $1/T_{\mathrm{1d}}^{\ast}$ versus $T$ in powder samples~7a
and~6a in Figs.~\ref{fig:T1d}(b) and~(c), respectively, show an almost
field independent peak, similar to the peaks in $1/T_1^\ast$ versus $T$ in
Figs.~\ref{fig:IT1beta}(b) and (c).  As discussed above, the peaks are
attributed to spin freezing of the magnetic defects.  For the single
crystals, we only extract $1/T_{\mathrm{1d}}^*$ values above 1.3~K for
reasons discussed above.  Here $1/T_{\mathrm{1d}}^*$ is nearly constant
from 4.2~K down to about 2~K, but then shows a decrease upon further
decrease in $T$.  This behavior is very different from that of the
powders.

The above root exponential relaxation behavior has been reported
previously in systems where the nuclear spin-lattice relaxation rate is
proportional to $1/r^6$, where $r$ is the distance between a nucleus and
a nearby paramagnetic center, and no nuclear spin diffusion takes
place.\cite{McHenry1972,Tse1968}  Nuclear spin-lattice relaxation due to
fluctuations of both dipolar and RKKY interactions have such $1/r^6$
dependences.  In general, one can write the nuclear spin-lattice
relaxation rate at the  nuclear site $\mathbf{r}$ due to a nearby
paramagnetic center at location $\mathbf{R}_{l}$ as 
\begin{equation} 
{1\over T_{1}\left(\mathbf{r,R}_{l}\right)}=C_{l} \frac{f(\theta
)}{\overline{f}}\frac{1}{\left\vert\mathbf{r-R}_{l}\right\vert^{6}},
\label{eqn:C}
\end{equation}
where $\mathbf{r-R}_{l}$ is the vector connecting the
paramagnetic center and nuclear spin, $\theta $ is the angle between
$\mathbf{r-R}_{l}$ and the external magnetic field,
$f(\theta)$ is the angular dependence of $1/
T_{1}(\mathbf{r,R}_{l})$, $\overline{f}$ is the average of 
$f(\theta )$ over all directions, and 
\begin{equation} 
C_{l} = C_{0}\frac{\tau _{l}}{1+\left( \omega
_{\rm{n}}\tau_{l}\right)^{2}}
\label{EqCl}
\end{equation} 
is a parameter proportional to the spectral density of
spin fluctuations (with correlation time $\tau_l$) at the nuclear Larmor
frequency $\omega _{\rm{n}}$.\cite{McHenry1972}  The defect contribution
of the recovery of the nuclear magnetization towards equilibrium is then
given as 
\begin{equation} 
p\left( \mathbf{r},t\right) = \exp \left[
-t\sum_{l}{1\over T_{1}\left( \mathbf{r,R}_{l}\right)}\right],
\end{equation}
where the sum is over all defect sites.  We first ignore
the spatial variation of the microscopic relaxation rate, i.e.\ we assume
that $\tau _{l}=\tau$ is the same for all defects, and concentrate on the
geometric inhomogeneity as caused by a varying distance between nuclear
spins and defects.  In order to evaluate the average over defect
positions we write
$\sum_{l}h( \mathbf{R}_{l}) =\sum_{i}g_{i}h( \mathbf{x}
_{i})$ where $h(\mathbf{y}_{i})$ is an arbitrary function of position
$\mathbf{y}_{i}$, $\mathbf{x}_{i}$ is a vanadium site position, we
assume the magnetic defects are at vanadium lattice sites, and the sum
over $i$ now runs over all vanadium sites, not only magnetic defect
sites.  The random variable $g_{i}$ is $1$ with probability
$n_{\rm defect}$ and $0$ with probability $\left( 1-n_{\rm
defect}\right)$ and assumed uncorrelated for different sites, i.e.\
defect positions are uncorrelated.  Following
Refs.~\onlinecite{McHenry1972} and
\onlinecite{Tse1968} we obtain in the limit of low defect concentration
($n_{\rm defect}\ll 1$) that 
\begin{equation} 
p\left( t\right) =\exp \left\{ -n_{\rm
defect}\sum_{i}\left[ 1-\exp \left(
{-t\over T_{1}\left(\mathbf{r,x}_{i}\right)}
\right) \right] \right\}.
\end{equation}
The sum over the lattice is evaluated as an integral.  In
the long time limit we obtain the result $p\left( t\right) $ as given in
Eq.~(\ref{eqn:rootexp}) where $1/T_{1\mathrm{d}}^{\ast }$ is then
given by\cite{Tse1968,McHenry1972}
\begin{equation} 
{1\over T_{1\mathrm{d}}^{\ast }} =\frac{16\pi ^{3}}{9}(\rho
_{N}n_{\mathrm{defect}})^{2}\frac{C_{0}\tau}{1+\left( \omega
_{\rm{n}}\tau\right) ^{2}},
\label{eqn:T1d}
\end{equation}
where $\rho _{N}$ is the number density of LiV$_{2}$O$_{4}$ formula units.

In the Appendix, we show that instead of solving for the relaxation curve,
we can understand the occurrence of a root exponential relaxation as
arising from our calculated probability distribution of nuclear $1/T_1$
values.  We will discuss the temperature and field dependences of
$1/T_{1\mathrm{d}}^{\ast }$ when we study the dynamics of the magnetic
defects in Sec.~\ref{sec:dynamicsI}.

\subsubsection{Hole Burning Experiment and the Dependence of Relaxation on
the Position in the Spectrum}

Bloembergen and coworkers\cite{Bloembergen1959} have considered the
problem of spin diffusion in the frequency domain (spectral diffusion) in
a spectrum with the same kind of inhomogeneous broadening as discussed
for the longitudinal spin relaxation.  The time for a hole to diffuse
through the whole spectrum by two-spin mutual spin flip is estimated to
be $T_{2}^{4}/T_{2}^{\ast 3}$, where $T_{2}$ is the intrinsic nuclear
spin-spin relaxation time and $T_{2}^{\ast}$ is the half width at half
maximum of the transient echo signal.  In the powder sample~6a
($n_{\mathrm{defect}}=0.83$~mol\%), $T_{2}\approx 200~\mathrm{\mu s}$
and $T_{2}^{\ast }\approx 5~\mathrm{\mu s}$, so $T_{2}^{4}/T_{2}^{\ast
3}=32$~s.  In the powder sample~6b ($n_{\mathrm{defect}}=0.21$~mol\%),
$T_{2}\approx 200~\mathrm{\mu s}$ and $T_{2}^{\ast }\approx
20~\mathrm{\mu s}$ which give $T_{2}^{4}/T_{2}^{\ast 3}=200$~ms.  Both
diffusion times are much longer than the values of $T_{1}^{\ast }$ at
4.2~K in each sample in Figs.~\ref{fig:IT1beta} and
\ref{fig:IT1beta-xstal}, and are thus consistent with the lack of
spectral diffusion in Fig.~\ref{fig:holerecovery}.

The higher relaxation rates at the wings of the spectrum compared to that
at the peak of the spectrum as shown in Fig.~\ref{fig:rlxSD7a} can also
be qualitatively explained within the approach where we only include the
geometric distribution of the nuclear spin to defect separations.  For
concreteness of discussion, we assume that the local field is purely
dipolar.  Denote the angle between the applied magnetic field and the
direction from a magnetic defect to a nuclear spin by $\theta $ and the
distance between the defect and the nuclear spin by $r$.  The NMR
frequency shift depends on $\theta $ and $r$ through $(1-3\cos ^{2}\theta
)/r^{3}$, while the nuclear spin-lattice relaxation rate depends on
$\theta $ and $r$ through $\sin^{2}\theta \cos^{2}\theta
/r^{6}$.\cite{Abragam}  The higher relaxation rates observed at the wings
compared to that at the peak of the spectrum is due to the monotonic
decrease of both the frequency shift and the nuclear spin-lattice
relaxation rates with increasing distance $r$.  The nuclear spins with
larger frequency shift will also have a higher probability of having
larger $1/T_{1}$ values.

\subsubsection{$^7$Li Nuclear Spin Diffusion\label{Sec:SpinDiffusion}}

\begin{figure}[tbp]
\centering
\includegraphics[width=3in]{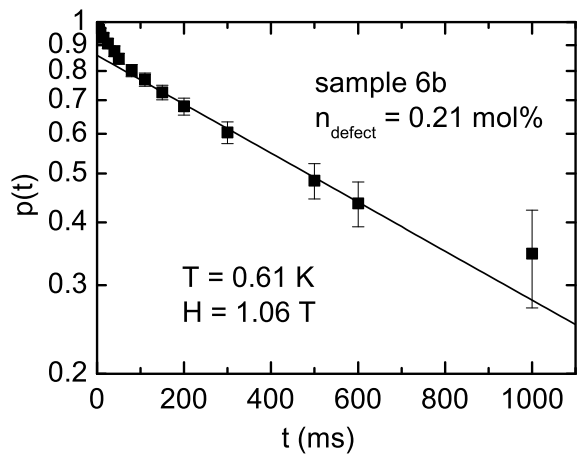}
\caption{Semilog plot of the nuclear spin relaxation function $p(t)$ in
Eq.~(\protect\ref{eqn:pt}) versus time $t$ after saturation for
LiV$_{2}$O$_{4}$ powder sample~6b with $n_{\mathrm{defect}}=0.21$~mol\% at
$H = 1.06$~T and
$T=0.61$~K\@.  The straight line is a single exponential fit to the data
at $t\ge 110$~ms.}
\label{fig:ptsd62061K}
\end{figure}

The $p(t)$ of powder sample~6b ($n_{\mathrm{defect}}=0.21$~mol\%) in
Fig.~\ref{fig:ptall}(a) deviates from a root exponential decay at
$t\gtrsim 100$~ms at $T=0.61$~K\@.  This can be attributed to the effect
of spin diffusion.\cite{Blumberg1960}  Spin diffusion tries to establish a
common spin temperature (i.e., the same longitudinal magnetization) among
nuclear spins at different distances from the defects and results in a
single exponential relaxation at long $t$.  Figure~\ref{fig:ptsd62061K}
displays $p(t)$ versus $t$ of the same data as in Fig.~\ref{fig:ptall}(a)
at $T=0.61$~K, but on a semilog scale, which suggests a single
exponential decay at $t\gtrsim 100$~ms.  A fit by $p(t)=A\exp(-t/T_1)$
to the data at $t\ge 110$~ms gives $1/T_1 = 1.1$~s$^{-1}$ and $A=0.86$.
The best fit is shown as the straight line in Fig.~\ref{fig:ptsd62061K}.

A crossover from a root exponential to a single exponential decay occurs
in the case of diffusion limited relaxation as discussed first by
Blumberg in Ref.~\onlinecite{Blumberg1960}.  The time $t_{\mathrm{c}}$, at
which the crossover from a root exponential to a single exponential decay
takes place, is related to the spin diffusion constant $D$
through\cite{Blumberg1960} 
\begin{equation} 
t_{\mathrm{c}}=C^{1/2}D^{-3/2},
\label{eqn:tc2}
\end{equation} 
where we dropped the subscript of $C_l$ in Eq.~(\ref{EqCl}) since we
assume the same spin dynamics for all the magnetic defects.  The diffusion
constant $D$ is related to the rate $W$ of mutual flips of nearest
neighbor nuclear spins through\cite{Abragam} 
\begin{equation} 
D = W a^{2},  
\label{eqn:D}
\end{equation} 
where $a$ is the distance between the two spins.  The rate
of the single exponential decay at long times in
Fig.~\ref{fig:ptsd62061K} is given by\cite{Blumberg1960} 
\begin{equation}
\frac{1}{T_1} = 8.5 \rho_Nn_{\mathrm{defect}}C^{1/4}D^{3/4},
\label{eqn:longT1}
\end{equation} 
where $\rho_N$ is the number density of LiV$_2$O$_4$ formula units.

In order to confirm the spin diffusion interpretation, below we will show
that the estimated crossover time $t_{\mathrm{c}}$ and $1/T_1$ are of the
same order of magnitude as the observed $t_{\mathrm{c}}\sim100$~ms and
$1/T_1=1.1$~s$^{-1}$, respectively.  The mutual spin-flip is due to
nuclear dipolar interactions and the value of $W$ can be estimated using
Fermi's golden rule.  For nuclear spins having $I=1/2$, after averaging
over the angular dependence, one obtains\cite{Abragam} 
\begin{equation}
W=\frac{2}{5}\pi\frac{\gamma_{\mathrm{n}}^{4}\hbar^{2}}{4a^{6}}\rho(0),
\label{eqn:W}
\end{equation} 
where $\rho(0)$ is the spectral density of the two spin
system at zero Zeeman energy and $\gamma_{\mathrm{n}}$ is the
gyromagnetic ratio of the nuclear spins.  The $^7$Li nuclei have spin
$I=3/2$, but an expression for $W$ when $I=3/2$ is not available, and the
above equation for $W$ should provide at least a rough estimate of $W$.
Approximating $\rho(0)$ by $1/\sqrt{2\pi
\langle \Delta \omega^{2} \rangle }$,\cite{Abragam} where $\langle \Delta
\omega^{2} \rangle=288$~kHz$^2$ is the Van~Vleck second moment of the
$^7$Li nuclei,\cite{Onoda1997} and taking $a = 3.57$ \AA, which is the
nearest-neighbor $^7$Li-$^7$Li distance in LiV$_2$O$_4$, we find $W=46\,
\mathrm{s}^{-1}$ from Eq.~(\ref{eqn:W}) and $D=5.9\times10^{-14}\mathrm{
cm^{2}/s}$ from Eq.~(\ref{eqn:D}).

The value of $C$ in Eq.~(\ref{eqn:tc2}) can be obtained from
Eq.~(\ref{eqn:T1d}) where $1/T_{1\mathrm{d}}^{\ast}$ is measured using
Eq.~(\ref{eqn:rootexp}) from the initial root exponential part of
$p(t)$ in Fig.~\ref{fig:ptall}(a) for powder sample~6b.  At $T=0.61$~K
and $H=1.06$~T, one obtains $1/T_{1\mathrm{d}}^{\ast} = 0.7$~s$^{-1}$, so
one has $C=1.4\times10^{-41}$~cm$^6$~s$^{-1}$.  Using the above
$D=5.9\times10^{-14}~\mathrm{cm^{2}/s}$ and
$n_{\mathrm{defect}} = 0.21$~mol\%, Eq.~(\ref{eqn:tc2}) yields the
crossover time $t_{\mathrm{c}}=220$~ms and Eq.~(\ref{eqn:longT1}) yields
the long time decay rate $1/T_1=1.9$~s$^{-1}$.  Due to the uncertainty in
our estimate of the parameter $D$ and the approximate nature of
Eq.~(\ref{eqn:tc2}), the estimated
$t_{\mathrm{c}}$ and $1/T_1$ values should be considered to be consistent
with the observed $t_{\mathrm{c}}\sim 100$~ms and $1/T_1=1.1$~s$^{-1}$,
respectively.

The absence of a deviation from root exponential behavior in samples~7a
($n_{\mathrm{defect}}=0.49$~mol\%) and~6a
($n_{\mathrm{defect}}=0.83$~mol\%) as shown in Figs.~\ref{fig:ptall}(b)
and~(c) may be due to the effect of inhomogeneous broadening, which
decreases the probability of overlap of Zeeman level splittings of
neighboring $^7$Li nuclei and results in a decrease in the spin diffusion
constant $D$.  Furthermore, due to the higher concentrations of the
defects, values of $p(t)$ at $t\gtrsim 100$~ms in these two samples are
much smaller than in the 0.21~mol\% sample, making such a deviation more
difficult to observe.

\subsubsection{\label{sec:dynamicsI}Magnetic Defect Spin Dynamics in the
Powder Samples}

In this section, we discuss the relation of the nuclear spin-lattice
relaxation rate to the dynamics of the magnetic defects in the powder
samples. We consider the weak collision limit $h\ll H$, where $h$ is the
magnitude of the local fluctuating field at the nuclear site and $H$ is
the magnitude of the static applied field.  The nuclear spin-lattice
relaxation rate $1/T_1$ due to an electronic magnetic defect spin at the
origin is then given by\cite{Abragam} 
\begin{equation}
\frac{1}{T_1}(\bm{r}) = \frac{1}{\hbar^2}\sum_{\alpha=x,y,z}
A_\alpha^2(\bm{r}) \int^{\infty}_{-\infty} \langle S_{\alpha}(0)
S_{\alpha}(t)\rangle \exp(i\omega_{\mathrm{n}}t) dt,
\end{equation} 
where $\bm{r}$ is the position of the nuclear spin with
respect to the magnetic defect, $A_\alpha(\bm{r})$ is the hyperfine
coupling constant between the nuclear spin and the magnetic defect,
$\omega_{\mathrm{n}}=\gamma_{\mathrm{Li}}H$ is the nuclear Larmor angular
frequency, and $\langle S_{\alpha}(0) S_{\alpha}(t)\rangle$ ($\alpha=x, y,
z$) are the magnetic defect spin autocorrelation functions.

As indicated in the ac susceptibility measurements, the peaks in
$1/T_{\mathrm{1d}}^*$ versus $T$ in Fig.~\ref{fig:T1d} are related to spin
freezing of the magnetic defects.  As a first attempt, we assume a single
exponential decay for the magnetic defect spin autocorrelation functions
and assume that the freezing process is due to an energy barrier so that
the correlation time $\tau$ follows 
\begin{equation}
\tau = \tau_0\exp\left(\frac{\Delta}{T}\right),
\label{eqn:acttau}
\end{equation} 
where $\tau_0$ is the fluctuation rate of the paramagnetic
defects at high temperature and $\Delta$ is the energy barrier in
temperature units.  For simplicity, we will assume that all the magnetic
defect spins in a sample have the same correlation time $\tau$.  For
dipolar or RKKY interactions, the nuclear spin-lattice relaxation rate of
a $^7$Li nucleus due to a nearby defect at distance $r$ is\cite{Abragam} 
\begin{equation}
\frac{1}{T_1}(r)= \frac{2R
\mu_{\mathrm{B}}^2\gamma_{\mathrm{Li}}^{2}S(S+1)}{5r^6}
\frac{\tau}{1+\omega_{\mathrm{n}}^2\tau^2},
\label{eqn:Lorentz1}
\end{equation} 
where the angular dependence is ignored and the prefactor
is written in such a way that $R=1$ would correspond to relaxation due
solely to the fluctuating dipolar field of the longitudinal component of
the magnetic defect spin.  The presence of additional relaxation channels
would increase the value of $R$.  By comparing Eqs. (\ref{eqn:C}),
(\ref{EqCl}), and (\ref{eqn:Lorentz1}), one has 
\[ C_0 = \frac{2}{5} R
\mu_{\mathrm{B}}^2\gamma_{\mathrm{Li}}^{2}S(S+1).
\] 
Inserting this expression for $C_0$ into Eq.~(\ref{eqn:T1d}), the measured
characteristic relaxation rate $1/T_{\mathrm{1d}}^*$ can be written as 
\begin{equation}
\frac{1}{T_{\mathrm{1d}}^*}= R
\frac{32\pi^3}{45}\mu_{\mathrm{B}}^2
\gamma_{\mathrm{Li}}^{2}S(S+1)\rho_N^2 n_{\mathrm{defect}}^2
\frac{\tau}{1+\omega_{\mathrm{n}}^2\tau^2}.
\label{eqn:Lorentz2}
\end{equation} 
At high temperatures, $\tau$ is generally much shorter than the inverse of
the nuclear Larmor frequency $1/\omega_{\mathrm{n}}$. As $\tau$ increases
with decreasing temperature $T$, a peak appears in $1/T_{\mathrm{1d}}^*$
versus $T$ at the temperature where $\tau = 1/\omega_{\mathrm{n}}$.

\begin{table}[tbp]
\caption{ Best fit values of prefactor $R$ and energy barrier $\Delta$
obtained by fitting the $1/T_{\mathrm{1\,d}}^*$ data in
Fig.~\protect\ref{fig:T1d} by a combination of
Eqs.~(\protect\ref{eqn:acttau}) and (\protect
\ref{eqn:Lorentz2}). In order to see the correlation between defect
concentrations and $R$ and $\Delta$, the values of
$n_{\mathrm{defect}}\protect\sqrt{S(S+1)}$ are also listed. }
\label{tbl:T1fit}
\begin{ruledtabular}
\begin{tabular}{cccc} 
Sample &  $n_{\rm defect} \sqrt{S(S+1)}$~(mol\%) &
$R$ & $\Delta$~(K)  \\ 
\hline 
6b &  0.85 & 0.04(4) & 1.1(1) \\ 
7a &  1.9 & 0.17(6) & 1.8(2) \\  
6a &  3.6  & 0.24(4) & 2.5(2) \\ 
\end{tabular}
\end{ruledtabular}
\end{table} 
We fit the $1/T_{\mathrm{1d}}^*$ data in Fig.~\ref{fig:T1d}
on all three powder samples simultaneously by the combination of
Eqs.~(\ref{eqn:acttau}) and (\ref{eqn:Lorentz2}).  Possible field and
temperature dependences of the parameter $R$ are ignored in the fit.
There are seven free parameters in the fit, $R$ and $\Delta$ for each
sample and $\tau_0$ which is assumed to be sample independent.  The
fitting results are displayed in Fig.~\ref{fig:T1d} by the solid curves.
The best fit value of $\tau_0$ is $4.1\times10^{-10}$~s and the best fit
values of $R$ and $\Delta$ for each sample are listed in
Table~\ref{tbl:T1fit}.  The energy barrier $\Delta$ increases with
increasing concentration of magnetic defects, which indicates that the
dynamic slowing down with decreasing temperature originates from the
interaction between the magnetic defects.  Interaction between magnetic
defects should increase with increasing concentration of the magnetic
defects since the average nearest-neighbor distance decreases.

The values of $R$ in all three samples are much less than unity, a fact
which cannot be explained by the presence of other nuclear spin-lattice
relaxation mechanisms since additional relaxation mechanisms would
increase $R$.  Such small values of $R$ might be related to the
spin-glass like freezing as observed in the ac magnetic susceptibility
measurements.  In spin glass systems, the spin autocorrelation functions
are highly nonexponential,\cite{Ogielski1985,Murani1981} which reduces
the spectral density of the magnetic defect spin fluctuations at $\omega
_{\mathrm{n}}$ as compared to the Lorentzian in Eq.~(\ref{eqn:Lorentz2}).
The reduction in spectral density thus results in a reduction in the
fitted value of $R$ in Eq.~(\ref{eqn:Lorentz2}).

\subsection{Effects of a Size Distribution of Magnetic 
Droplets\label{Sec:Droplets}}

Our previous discussion demonstrated that the non-exponential time
dependence of the $^7$Li nuclear spin-lattice relaxation in the
sample of single crystals below 1.3~K cannot be understood in terms
of the interaction of the nuclear spins with separable contributions from 
a heavy Fermi liquid and a random distribution of point-like
magnetic defects.  This problem is exemplified by the unphysical
$p(t)$ data at long times in Fig.~\ref{fig:ptxtal}(a) where the data
move away from equilibrium rather than towards
equilibrium with increasing time $t$.  An appealing approach
that refines our previous analysis is based on the hypothesis that the
magnetic defects are not point-like, but are rather magnetic droplets
with an average size significantly larger than an atomic size.  This
hypothesis is supported by the large average spins of the magnetic
defects inferred above in Sec.~\ref{Sec:ndefect} and
previously\cite{Kondo1999, Das2006} to be $S\sim 2$--4.  We envision that
the magnetic droplets have a variable size and spin and that the
corresponding microscopic internal relaxation time $\tau$ varies with
droplet size.  Our arguments parallel those for the unusual spin dynamics
due to statistically rare fluctuations that becomes crucial in the
context of Griffiths singularities close to phase
transitions.\cite{Griffiths,Bray87} 

In most three-dimensional 
systems a high temperature Curie-Weiss behavior of the susceptibility
with Weiss temperature $\theta$ is indicative for magnetic ordering at a
lower temperature $T \sim \theta$.  However, the geometric frustration
for antiferromagnetic ordering within the vanadium sublattice of the
spinel structure is likely the reason why long-range
antiferromagnetic order is suppressed in
pure \textrm{LiV}$_{2}$\textrm{O}$_{4}$ and a heavy electron state emerges
instead.  Defects in the crystal structure can locally lift the
frustration and easily cause magnetic order in a finite region of volume
$\simeq \xi^{d}$ around the crystal defect where the linear size $\xi$
is, due to the proximity to an ordered state, expected to be larger than
the interatomic spacing.  Here $d$ is the dimensionality of the
system.  These finite regions of magnetic order are what we are calling
magnetic droplets.

Fluctuations in the local tendency
towards order usually lead to a probability density for the linear droplet
size $\xi$ that decreases exponentially with the volume of the
droplet.  In three dimensions one obtains\cite{Thill1995}
\begin{equation} 
P\left( \xi \right) ={3\xi ^{2}\over \xi _{0}^{3}}\exp \left[ -\left( \xi
/\xi _{0}\right) ^{3}\right],
\end{equation} 
where $\xi_{0}$ is the mean droplet
size.  It is then natural to assume that the typical internal excitation
energy of a droplet decreases with increasing droplet size.  These 
excitations can change the magnitude and/or
direction of the magnetic moment of the droplet.  Often the
excitation energy $\varepsilon \sim \hbar/\tau$ varies with $\xi$
according to a power law 
\begin{equation}
\tau ^{-1}=D\xi ^{-\psi },
\label{power}
\end{equation}
where we have set $\hbar = 1$ and $\tau$ is the relaxation time of
the excitation.  The arguments of
Ref.~\onlinecite{Bray87} yield a result $\psi =d$ for classical Heisenberg
spins, where $d = 3$ for the present problem.  Quantum effects can then
yield deviations from this behavior and typically yield exponents $\psi$
that are larger than the above classical result.\cite{Vojta05} In some
cases, such as Ising spins in a magnetic field or Heisenberg spins in a
metallic host,\cite{Vojta05} the quantum dynamics can lead to a droplet
dynamics in three dimensions where 
\begin{equation}
\tau ^{-1}\propto e^{-\alpha \left( \xi /\xi _{0}\right) ^{3}}. 
\label{exp}
\end{equation}
Finally, in case of magnetic defects with Ising anisotropy inside a
metallic host, the quantum dynamics of the defect leads to a freezing of
all droplets beyond a certain size, typically of order $\xi _{0}$ (see
Refs.~\onlinecite{Millis01} and~\onlinecite{Millis05}).

To illustrate the effects of such droplet dynamics we concentrate first
on the case Eq.~(\ref{power}).  In the absence of a microscopic model of
the droplet spin dynamics we leave $\psi $ an open parameter of the
model.  We start from Eqs.~(\ref{eqn:rootexp}) and (\ref{eqn:T1d}) and
write
\begin{equation} 
p\left( t\right) =\int d\xi \,P(\xi) \,\exp
\left[-\left( \frac{t}{T_{1}^{\ast}\left(\xi\right) }\right)
^{1/2}\right],
\label{Eqpave}
\end{equation}
where $T_{1}^{\ast }\left( \xi \right) ^{-1}$ is given in
Eq.~(\ref{eqn:Lorentz2}) with $\tau$ replaced by $\tau \left(\xi
\right)$. In the evaluation of this average we have two very distinct
limits.  If $\omega _{\rm n}\tau \ll 1$ the long time behavior of the
system is dominated by small clusters since $T_{1}^{\ast -1}\propto \tau
$ and slow nuclear relaxation is caused by fast droplet dynamics.   In
this limit, the average over droplet sizes will not cause any changes in
the stretched exponential behavior as compared to our previous, purely
geometric, considerations for point-like magnetic defects.  The inequality
$\omega _{\rm n}\tau \ll 1$ is expected to be valid at higher
temperatures, before a freezing or dramatic slowing down of the droplet
moments sets in with decreasing temperature.  This is fully consistent
with our findings that we obtain $\beta =\frac{1}{2}$ in this regime.  The
situation changes dramatically in the limit where $\omega _{\rm n}\tau
\gg 1$, relevant at lower temperature, i.e.\ most likely for $T\lesssim
1.5~\mathrm{K}$.  Now
$T_{1}^{\ast -1}\propto\tau ^{-1}$ and slow nuclear relaxation is tied to
slow droplet relaxation.  In this regime we find 
\begin{equation}
\frac{1}{T_{1}^{\ast }\left( \xi \right) }\ \propto \xi ^{-\psi }.
\end{equation} 
In the long time limit, the average over the droplet sizes in
Eq.~(\ref{Eqpave}) can be performed via saddle point integration and
yields 
\begin{equation} 
p\left( t\right) =\exp \left[ -\left(
\frac{t}{T_{1,\mathrm{droplet}}^{\ast }}
\right) ^{\beta }\right]
\end{equation} 
with 
\begin{equation}
\beta =\frac{3}{6+\psi }\ ,
\end{equation}
i.e., the additional inclusion of droplet size variations
yields a stretched exponent $\beta <\frac{1}{2}$.  The characteristic
relaxation rate of the droplets is
\begin{equation} 
{1\over T_{1,\mathrm{droplet}}^{\ast}}\simeq R\frac{32\pi^3}{45}\frac{\mu
_{\mathrm{B}}^{2}\gamma _{
\mathrm{Li}}^{2}S(S+1)\rho _{N}^{2}n_{\mathrm{defect}}^{2}}{\omega
_{\mathrm{ n}}^{2}\tau(\xi_{0}) }.
\end{equation}
We emphasize that the static size distribution of magnetic droplets only
becomes apparent by NMR at low temperatures such that $\omega_{\rm n}\tau
> 1$, where $\omega_{\rm n}$ is the nuclear Larmor frequency.  We also
emphasize that the analyses in this section do not include interactions
between droplets and a possible resultant spin glass phase.  Above
$T\simeq 1.5~\mathrm{K}$ size variations of the droplets will not affect
the long time nuclear relaxation, even if they are present.

In case of an exponential dependence of the droplet excitation energy,
Eq.~(\ref{exp}), the long time dynamics of the nuclear spins at low $T$
is even more dramatically affected and changes to a power law decay
$p\left( t\right) \propto t^{-\lambda }$, with nonuniversal exponent
$\lambda$.  At the same time heat capacity and susceptibility
measurements also experience power law behavior.\cite{Vojta05} In this
regime droplet quantum fluctuations will not only dominate the long time
dynamics of the nuclear decay, but also thermodynamics quantities.  This
might be responsible for the fact that it is not possible any longer to
clearly separate the response of the underlying Fermi liquid from that of
the droplet at low temperatures in our single crystals, as indicated by
Fig.~\ref{fig:ptxtal}(a).  

Our data for single crystals do not allow at present to distinguish
between the different scenarios outlined in this section.  They do however
suggest that dynamics of the magnetic droplets plays an important role.

\section{\label{scn:conclusions}Summary and Conclusions}

The modeling approach in Sec.~\ref{sec:geometry} gives a good
description of our $^7$Li NMR results from 0.5 to 4.2~K for our
LiV$_{2}$O$_{4}$ powder samples containing magnetic defect concentrations
up to 0.83~mol\%.  This approach assumes that (i) there is a random
distribution of magnetic point defects, (ii) the heavy Fermi liquid in
magnetically pure LiV$_{2}$O$_{4}$ survives in samples containing up to
$\sim 0.8$~mol\% magnetic defects, and that (iii) the influences of the
magnetic defects and of the Fermi liquid on the magnetization and NMR
properties are separable.  This description explains very well the 
defect concentration-independent $\chi_0$ value from our low-temperature
magnetization measurements, the inhomogeneous broadening of the $^{7}$Li
NMR spectrum, the nonexponential $^7$Li nuclear spin-lattice relaxation
versus time behavior, and the lack of spectral diffusion in the $^7$Li
NMR hole burning experiments.  It also explains the smaller $^7$Li nuclear
spin-lattice relaxation rate at the peak of the spectrum as compared to
that at the wings.  However, it is hard to reconcile the picture of
magnetic point defects with the high magnetic moments for the defects
(spins of 2--4) deduced here (see Table~\ref{tbl:Mfit}) and in
Refs.~\onlinecite{Kondo1999} and
\onlinecite{Das2007} from magnetization measurements.  These large defect
spin values suggest that the magnetic defects may not behave like
point-like magnetic moments under all circumstances.  In
Sec.~\ref{Sec:Droplets} we discussed magnetic defects that are more
extended entities that we have called magnetic droplets with a
distribution of sizes, and likely a distribution of spin values.  We
showed that such a size distribution can affect the NMR magnetization
recovery at long times at low temperatures in the regime where
interactions between the magnetic droplets can be neglected.

Our study shows that there can be different kinds of magnetic defects in
the LiV$_2$O$_4$ system.  As revealed by the nuclear spin lattice
relaxation rate data and ac magnetic susceptibility measurements at
14~MHz, it is amazing that the magnetic defects in the powder samples
undergo a spin glass-like freezing below 1~K, while the magnetic defects
in the single crystals with a similar magnetic defect concentration
exhibit a very different behavior at such low temperatures, with no
evidence for spin freezing.   The different kinds of magnetic defects
and/or interactions in the crystals and powders must be associated with
different types of structural defects in the system, which might be
expected because the crystals are grown at about 1000~$^\circ$C whereas
the powders are synthesized at 700$^\circ$C\@.  Different types of
magnetic defects were even found in an annealing study of different single
crystals,\cite{Das2007} where heat treatment at 700~$^\circ$C was found
to remove the magnetic defects in one but not in other single crystals.

Further experiments on the single crystals are urgently needed at $\sim
1$~K and below.  These experiments should test whether
the Fermi liquid is modified by quantum fluctuations of large magnetic
droplets in the single crystals at $T<1.3$~K, whether the
magnetic properties of the crystals contain separable contributions from
the Fermi liquid and the magnetic defects, and whether the
conduction electrons in the crystals even form a Fermi liquid.  In
addition, the origin of the distinct kinks at about 1.4~K in the
temperature dependences of $1/T_1^*$ and $\beta$ in Fig.~\ref{fig:IT1beta}
and of $M(\infty)T$ in Fig.~\ref{fig:IT1beta-xstal} for the single
crystals remains to be explained.

\begin{acknowledgments} We are grateful to B. J. Suh for many beneficial
suggestions.  Work at the Ames Laboratory was supported by the Department
of Energy-Basic Energy Sciences under Contract No.~DE-AC02-07CH11358.
\end{acknowledgments}

\appendix*
\section{Probability Distribution Underlying Stretched Exponential
Relaxation with ${\mathbf{\beta = 1/2}}$}

Here we discuss the stretched exponential $^7$Li nuclear relaxation versus
time following saturation that arises from interactions between the
$^7$Li nuclear spins and a low concentration of magnetic defect spins,
which is $p(t)$ in Eq.~(\ref{eqn:rootexp}).  We demonstrate that instead
of theoretically solving for the relaxation curve and showing that it is a
stretched exponential with $\beta = 1/2$ as in Sec.~\ref{Sec:NSLR}, we can
understand the occurrence of this root exponential relaxation as arising
from the probability distribution of nuclear $1/T_1$ values.

\begin{figure}[tbp]
\centering\includegraphics[width=3in]{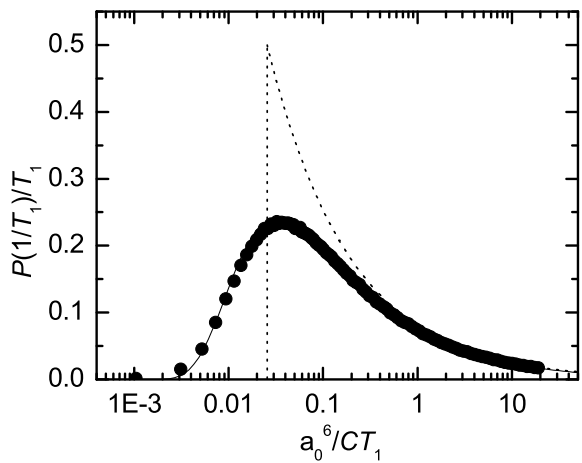}
\caption{$^7$Li nuclear spin-lattice relaxation rate $1/T_1$ probability
distributions $P(1/T_1)$, normalized by $T_1$, due to dilute point-like
paramagnetic defects.  The filled circles are results obtained from
computer simulation of LiV$_{2}$O$_{4}$ where the magnetic defects
randomly occupy vanadium sites with probability 0.25\%.  The solid curve
is the best fit of these data by Eq.~(\protect\ref{eqn:pbeta12}) with
$1/T_1^*=0.067(1)$~s$^{-1}$.  The dotted curve is a plot of $p(x) =
0.8x^{-3/2}$ with a small $x$ cutoff at
$x_c = 0.0256$, where $x \equiv a_0^6/CT_1$.  The lower cutoff $x_c$ is
chosen so that the normalization condition $\protect\int_{x_c}^\infty
P(x) \,dx =1$ is satisfied.}
\label{fig:T1distb0p5}
\end{figure}

We assume that $p(t)$ in Eq.~(\ref{eqn:rootexp}) is due to a continuous
sum of exponential decays with a distribution of relaxation rates
$1/T_1$.  Then one can write the stretched exponential relaxation
function in Eq.~(\ref{eqn:rootexp}) as 
\begin{equation}
e^{-(t/T_{\mathrm{1d}}^\ast)^\beta} = \int_0^\infty 
P(s,\beta) e^{-st/T_{\mathrm{1d}}^\ast}ds,
\label{EqLaplace}
\end{equation}
where $s$ equals $1/T_1$ normalized by $1/T_{\mathrm{1d}}^{\ast}$, i.e.,
$s\equiv$ $T_{\mathrm{1d}}^{\ast}/T_{1}$, and $P(s,\beta)$ is the
probability density for occurrence of $s$ for a fixed exponent
$\beta$ with $0 < \beta \leq 1$.  Thus the stretched exponential function
is the Laplace transform of $P(s,\beta)$.  Closed analytic expressions for
$P(s,\beta)$ with rational values of $\beta$ can be obtained from the
inverse Laplace transform of the stretched exponential
function, and physical interpretations of the
parameters $1/T_{\mathrm{1d}}^*$ and $\beta$ have been
determined.\cite{Johnston2005, Johnston2006} We show below 
that the probability distribution of $1/T_1$ due to dipolar interaction
of nuclear spins with dilute magnetic defects corresponds very well to the
$1/T_1$ distribution leading to the stretched exponential
relaxation in Eqs.~(\ref{eqn:rootexp}) and (\ref{EqLaplace}) with
$\beta=1/2$.  This probability density is \cite{Johnston2005,
Johnston2006} 
\begin{equation} 
P(s,1/2)=\frac{e^{-\frac{1}{4s}}}{\sqrt{4\pi}s^{3/2}}.
\label{eqn:pbeta12}
\end{equation} 
A plot of this distribution function is given below as the solid curve in
Fig.~\ref{fig:T1distb0p5}.  $P(s,1/2)$ is proportional to
$s^{-3/2}$ for large $s$ [in general $P(s,\beta)$ at large $s$ is
proportional to $s^{-(1 + \beta)}$], and has a low-$s$ cutoff since
$e^{-1/4s}$ exponentially approaches zero at small $s$ values (also
true in general for arbitrary $\beta$).\cite{Johnston2006}

A qualitative $s^{-3/2}$ dependence of the $1/T_1$
probability distribution arises due to a $r^{-6}$ dependence of $1/T_1$ as
in Eq.\ (\ref{eqn:C}) as follows.  Ignoring the angular dependence
in Eq.~(\ref{eqn:C}) and assuming the same dynamics for all the
magnetic defect spins such that $C_l=C$, in the single paramagnetic center
limit the distribution of $s$ arising from a \emph{continuum} distribution
of nuclear spins around a magnetic defect is\cite{Johnston2005} 
\begin{equation} 
P_{\mathrm{geo}}(s) \propto
r^{2}\frac{\mathrm{d}r}{\mathrm{d}s}\Big|_{r=
\big(\frac{CT_{\mathrm{1d}}^*}{s}\big)^{1/6}}
\propto s^{-3/2}.
\label{eqn:psnglm}
\end{equation} 
This $s$ dependence is the same as the large-$s$ limit of the result
in Eq.~(\ref{eqn:pbeta12}) for stretching exponent $\beta = 1/2$ noted
above.  The distribution (\ref{eqn:psnglm}) diverges as $1/T_1$
approaches zero.  This divergence is caused by the single impurity
approximation.  Nuclei with $1/T_{1}$ approaching zero correspond to
those far away from the paramagnetic center.  Due to the finite distance
between different paramagnetic centers in a system with a finite
concentration of them, the actual probability of finding a nuclear spin
with $1/T_{1}\rightarrow0$ should instead vanish, so a low $1/T_1$ cutoff
has to be applied.  This results in a distribution function with a shape
(see the dashed curve in Fig.~\ref{fig:T1distb0p5} below) roughly similar
to that in  Eq.~(\ref{eqn:pbeta12}) (the solid curve in
Fig.~\ref{fig:T1distb0p5} below).

We have carried out a numerical simulation of the $1/T_1$
probability distribution that turns out to be in very good agreement with
the exact result for the probability distribution in
Eq.~(\ref{eqn:pbeta12}) for the stretched exponential function with
$\beta = 1/2$.  In the simulation, we calculated the $1/T_1$ distribution
of $^7$Li nuclei due to a random distribution of dilute
point-like paramagnetic centers (defects) in the LiV$_2$O$_4$
spinel structure.  The paramagnetic defects randomly occupy the vanadium
sites with a probability of $0.25\%$ ($n_{\mathrm{defect}}=0.5$~mol\%)
and the configuration of the random defects repeats every 80 unit cells
in all three crystallographic axis directions.  The $1/T_1$ of each $^7$Li
nucleus is calculated using\cite{Abragam} 
\begin{equation}
\frac{1}{T_1} = C \sum_i
\frac{15\sin^2\theta_i\cos^2\theta_i}{2}\frac{1}{r_i^6}, 
\label{eqn:T1sim}
\end{equation} 
where $C$ is a constant, $r_i$ is the distance between paramagnetic center
$i$ and the $^7$Li nucleus, and $\theta_i$ is the angle between the
applied magnetic field and the vector from paramagnetic center $i$ to the
$^7$Li nucleus.  The applied magnetic field was arbitrarily chosen to be
along the $\langle001\rangle$ direction.  Equation (\ref{eqn:T1sim}) is
the nuclear spin-lattice relaxation due to the dipolar magnetic field
fluctuations from the \emph{longitudinal} spin component of the
paramagnetic defects.\cite{Abragam}  In the presence of a strong applied
magnetic field, the \emph{transverse} spin fluctuation is often modulated
by the Larmor frequency of the \emph{electronic} spins and thus has
negligible contribution to $1/T_1$ for which magnetic field fluctuations
at the \emph{nuclear} Larmor frequency are most important.\cite{Abragam} 
The summation over $i$ in Eq.~(\ref{eqn:T1sim}) includes all
magnetic defects with $r_i < 20 a_0$, where $a_0=8.24$~\AA\ is the lattice
constant of cubic LiV$_2$O$_4$.  We checked that changing the summation
range to $r_i < 10 a_0$ resulted in a negligible change in the $1/T_1$
distribution.

The distribution of $1/T_1$ resulting from the above simulation is
displayed as the filled circles in Fig.~\ref{fig:T1distb0p5}, where the
distribution is normalized by $T_1$.  The maximum relaxation rate
plotted in Fig.~\ref{fig:T1distb0p5} at $a_0^6/CT_1 \approx 20$ is not
a large relaxation rate cutoff to the probability distribution.  Data
at larger relaxation rates are not plotted due to insufficient
statistics.  The simulated $1/T_1$ distribution in
Fig.~\ref{fig:T1distb0p5} can be fitted very well by
Eq.~(\ref{eqn:pbeta12}) with
$1/T_{\mathrm{1d}}^* = 0.067(1) C/a_0^6$, as shown by the solid curve in
Fig.~\ref{fig:T1distb0p5}. 
$1/T_{\mathrm{1d}}^*$ calculated from Eq.~(\ref{eqn:T1d}) is equal to
$0.088 C/a_0^6$, close to the simulated result.  The difference may be
due to the neglected angular dependence in deriving Eq.~(\ref{eqn:T1d}). 
For comparison, a simple power law distribution $P(x\equiv a_0^6/CT_1) =
0.08 x^{-3/2}$ with a small
$x$ cutoff of $x_c=0.0256$ is also displayed in
Fig.~\ref{fig:T1distb0p5}.  This $x$ dependence is the same as in
Eq.~(\ref{eqn:psnglm}) and is the same as the asymptotic large-$s$
dependence of Eq.~(\ref{eqn:pbeta12}).  The prefactor 0.08 is chosen to
make the distribution overlap with the simulated result at large $x$, and
the low-$x$ cutoff $x_c=0.0256$ is determined from the normalization
condition $\int_{x_c}^\infty P(x) \,dx =1$.


\begin{thebibliography}{99}

\bibitem{Kondo1997} S. Kondo, D. C. Johnston, C. A. Swenson, F. Borsa, A.
V. Mahajan, L. L. Miller, T. Gu, A. I. Goldman, M. B. Maple, D. A.
Gajewski, E. J. Freeman, N. R. Dilley, R. P. Dickey, J. Merrin, K.
Kojima, G. M. Luke, Y. J. Uemura, O. Chmaissem, and J. D. Jorgensen,
Phys.\ Rev.\ Lett.\ \textbf{78}, 3729 (1997).

\bibitem{Stewart1984} G. R. Stewart, Rev.\ Mod.\ Phys.\ \textbf{56}, 755
(1984).

\bibitem{Takagi1999} H. Takagi, C. Urano, S. Kondo, M. Nohara, Y. Ueda, T.
Shiraki, and T. Okubo, Mater.\ Sci.\ Eng.\ B \textbf{63}, 147 (1999).

\bibitem{Urano2000} C. Urano, M. Nohara, S. Kondo, F. Sakai, H. Takagi, T.
Shiraki, and T. Okubo, Phys.\ Rev.\ Lett.\ \textbf{85}, 1052 (2000).

\bibitem{Kadowaki1986} K. Kadowaki and W. B. Woods, Solid State
Commun.\ \textbf{58}, 507 (1986).

\bibitem{Fulde2004} P. Fulde, J. Phys.: Condens.\ Matter \textbf{16}, S591
(2004).

\bibitem{Arita2007} R. Arita, K. Held, A. V. Lukoyanov, and V. I.
Anisimov, Phys.\ Rev.\ Lett.\ \textbf{98}, 166402 (2007).

\bibitem{Yushankhai2007} V. Yushankhai, A. Yaresko, P. Fulde, and P.
Thalmeier, Phys.\ Rev.\ B \textbf{76}, 085111 (2007).

\bibitem{Mahajan1998} A. V. Mahajan, R. Sala, E. Lee, F. Borsa, S. Kondo,
and D. C. Johnston, Phys.\ Rev.\ B \textbf{57}, 8890 (1998).

\bibitem{Johnston1976} D. C. Johnston, J. Low Temp.\ Phys.\ \textbf{25},
145 (1976).

\bibitem{Dalton1994} M. Dalton, D. P. Tunstall, J. Todd, S. Arumugam, and
P. P. Edwards, J. Phys.: Condens.\ Matter \textbf{6}, 8859 (1994).

\bibitem{Johnston2005} D. C. Johnston, S.-H. Baek, X. Zong, F. Borsa, J.
Schmalian, and S. Kondo, Phys.\ Rev.\ Lett.\ \textbf{95}, 176408 (2005).

\bibitem{Kaps2001} H. Kaps, M. Brando, W. Trinkl, N. B\"uttgen, A. Loidl,
E.- W. Scheidt, M. Klemm, and S. Horn, J. Phys.: Condens.\ Matter
\textbf{13}, 8497 (2001).

\bibitem{Kondo1999} S. Kondo, D. C. Johnston, and L. L. Miller, Phys.\
Rev.\ B \textbf{59}, 2609 (1999).

\bibitem{Vannette2007} M. D. Vannette, A. Safa-Sefat, S. Jia, S. A. Law,
G. Lapertot, S. L. Bud'ko, P. C. Canfield, J. Schmalian, and R. Prozorov,
J. Mag.\ Mag.\ Mater.\ \textbf{320}, 354 (2007).

\bibitem{Das2006} S. Das, X. Ma, X. Zong, A. Niazi, and D. C. Johnston,
Phys.\ Rev.\ B \textbf{74}, 184417 (2006).

\bibitem{Das2007} S. Das, X. Zong, A. Niazi, A. Ellern, J. Q. Yan, and
D.~C. Johnston, Phys.\ Rev.\ B \textbf{76}, 054418 (2007).

\bibitem{Corson} D. Corson and P. Lorrain, \textit{Introduction to
Electromagnetic Fields and Waves} (W. H. Freeman and Company, San
Francisco, 1962).

\bibitem{Fukushima} E. Fukushima and S. B. W. Roeder, \textit{Experimental
Pulse NMR: A Nuts and Bolts Approach} (Perseus Books, Cambridge, 1981).

\bibitem{Abragam} A. Abragam, \textit{Principles of Nuclear Magnetism}
(Oxford University Press, Oxford, 1982).

\bibitem{Vleck1948} J. H. Van Vleck, Phys.\ Rev.\ \textbf{74}, 1168
(1948).

\bibitem{Drain1962} L. E. Drain, Proc.\ Phys.\ Soc.\ London \textbf{80},
1380 (1962).

\bibitem{Onoda1997} M. Onoda, H. Imai, Y. Amako, and H. Nagasawa, Phys.\
Rev.\ B \textbf{56}, 3760 (1997).

\bibitem{Slichter} C. P. Slichter, \textit{Principles of Magnetic
Resonance} (Springer, Berlin, 1990), 3rd ed.

\bibitem{Walstedt1974} R. E. Walstedt and L. R. Walker, Phys.\ Rev.\ B 
\textbf{9}, 4857 (1974).

\bibitem{Johnston2006} D. C. Johnston, Phys.\ Rev.\ B \textbf{74}, 184430
(2006).

\bibitem{McHenry1972} M. R. McHenry, B. G. Silbernagel, and J. H. Wernick,
Phys.\ Rev.\ B \textbf{5}, 2958 (1972).

\bibitem{Tse1968} D. Tse and S. R. Hartmann, Phys.\ Rev.\ Lett.\
\textbf{21}, 511 (1968).

\bibitem{Bloembergen1959} N. Bloembergen, S. Shapiro, P. S. Pershan, and
J. O. Artman, Phys.\ Rev.\ \textbf{114}, 445 (1959).

\bibitem{Blumberg1960} W. E. Blumberg, Phys.\ Rev.\ \textbf{119}, 79
(1960).

\bibitem{Ogielski1985} A. T. Ogielski, Phys.\ Rev.\ B \textbf{32}, 7384
(1985).

\bibitem{Murani1981} A. P. Murani, J. Magn.\ Magn.\ Mater.\ \textbf{22},
271 (1981).

\bibitem{Griffiths} R. B. Griffiths, Phys. Rev. Lett. \textbf{23}, 17
(1969).

\bibitem{Bray87} A. J. Bray, Phys.\ Rev.\ Lett.\ \textbf{59}, 586 (1987).

\bibitem{Thill1995} M. J. Thill and D. A. Huse, Physica~A
{\bf 214}, 321 (1995).

\bibitem{Vojta05} T. Vojta and J. Schmalian, Phys.\ Rev.\ B \textbf{72},
045438 (2005).

\bibitem{Millis01} A. J. Millis, D. K. Morr, and J. Schmalian, Phys.\
Rev.\ Lett.\ \textbf{87}, 167202 (2001).

\bibitem{Millis05} A. J. Millis, D. K. Morr, and J. Schmalian, Europhys.\
Lett.\ \textbf{72}, 1052 (2005).

\end{thebibliography}
\end{document}